\documentclass[sigconf]{acmart}
\usepackage[T1]{fontenc}
\usepackage[utf8]{inputenc}
\usepackage{amsmath}
\usepackage{xcolor}
\renewcommand\footnotetextcopyrightpermission[1]{} 
\usepackage{mathptmx} 

\newcommand{\ignore}[1]{}
\usepackage{fancyhdr}
\usepackage[normalem]{ulem}
\usepackage{microtype}
\usepackage{xspace}
\usepackage{physics}
\usepackage[]{subfig}
\usepackage{float}
\usepackage{textcomp}
\renewcommand{\textrightarrow}{$\rightarrow$}

\newcommand{\ibmnameshort} {{\em IBMQ14}}
\newcommand{\compilername} {TriQ\xspace}

\newcommand{\noopt} {TriQ-N\xspace}
\newcommand{\oneqopt} {TriQ-1QOpt\xspace}
\newcommand{\commopt} {TriQ-1QOptC\xspace}
\newcommand{\noiseopt} {TriQ-1QOptCN\xspace}
\newcommand{\Qiskit} {Qiskit\xspace}
\newcommand{\Quil} {Quil\xspace}

\newcommand{\syscnt} {seven\xspace}

\title{Full-Stack, Real-System Quantum Computer Studies: Architectural Comparisons and Design Insights}

\author{Prakash Murali}
\affiliation{%
  \institution{Princeton University}
}
\authornote{Prakash Murali is the corresponding author and can be reached at pmurali@cs.princeton.edu.}

\author{Norbert Matthias Linke}
\affiliation{%
  \institution{University of Maryland}
}
\author{Margaret Martonosi}
\affiliation{%
  \institution{Princeton University}
}
\author{Ali Javadi Abhari}
\affiliation{%
  \institution{IBM T. J. Watson Research Center}
}
\author{Nhung Hong Nguyen}
\affiliation{%
  \institution{University of Maryland}
}
\author{Cinthia Huerta Alderete}
\affiliation{\institution{University of Maryland}}
\affiliation{\institution{Instituto Nacional de Astrofísica, Óptica y Electrónica}}

\renewcommand{\shortauthors}{P. Murali et al.}
\setcopyright{none}

\begin{abstract}
In recent years, Quantum Computing (QC) has progressed to the point where small working prototypes are available for use. Termed Noisy Intermediate-Scale Quantum (NISQ) computers, these prototypes are too small for large benchmarks or even for Quantum Error Correction (QEC), but they do have sufficient resources to run small benchmarks, particularly if compiled with optimizations to make use of scarce qubits and limited operation counts and coherence times.  QC has not yet, however, settled on a particular preferred device implementation technology, and indeed different NISQ prototypes implement qubits with very different physical approaches and therefore widely-varying device and machine characteristics. 

Our work performs a full-stack, benchmark-driven hardware-software analysis of QC systems. We evaluate QC architectural possibilities, software-visible gates, and software optimizations to tackle fundamental design questions about gate set choices, communication topology, the factors affecting benchmark performance and compiler optimizations. In order to answer key cross-technology and cross-platform design questions, our work has built the first top-to-bottom toolflow to target different qubit device technologies, including superconducting and trapped ion qubits which are the current QC front-runners. We use our toolflow, \compilername, to conduct {\em real-system} measurements on \syscnt running QC prototypes from three different groups, IBM, Rigetti, and University of Maryland. Overall, we demonstrate that leveraging microarchitecture details in the compiler improves program success rate up to 28x on IBM (geomean 3x), 2.3x on Rigetti (geomean 1.45x), and 1.47x on UMDTI (geomean 1.17x), compared to vendor toolflows. In addition, from these real-system experiences at QC's hardware-software interface, we make observations and recommendations about native and software-visible gates for different QC technologies, as well as  communication topologies, and the value of noise-aware compilation even on lower-noise platforms. This is the largest cross-platform real-system QC study performed thus far; its results have the potential to inform both QC device and compiler design going forward.
\end{abstract}

\begin{document}
\maketitle
\thispagestyle{firstpage}
\pagestyle{plain}

\section{Introduction}
Quantum computing (QC) is emerging as a promising paradigm for solving classically intractable computational problems in areas such as machine learning \cite{quantum_ml1, quantum_ml2}, cryptography \cite{shor1}, chemistry \cite{vqe1, vqe2} and others. QC devices represent information using \emph{qubits} (quantum bits) and perform operations based on quantum mechanical principles such as superposition and entanglement to achieve speedups over classical algorithms. 

In recent years, QC implementations have advanced considerably.  QC prototypes with up to 16 qubits are available for broad public use \cite{ibmq} and larger 49-72 qubit systems are either announced or in use \cite{googlebristlecone,ibm50q,intelq}.  Much like the early days of classical (i.e. non-quantum) computing, however, QCs have not yet converged on a specific candidate device technology.  Front-runner technologies today include superconducting transmon qubits \cite{superc1, superc2} and trapped ion qubits \cite{trappedion1, trappedion2, trappedion3}, with other candidate technologies also of considerable interest \cite{majorana2, majorana3, spinqubits}. 

 \begin{figure*}[t]
 \centering
     \includegraphics[scale=0.35]{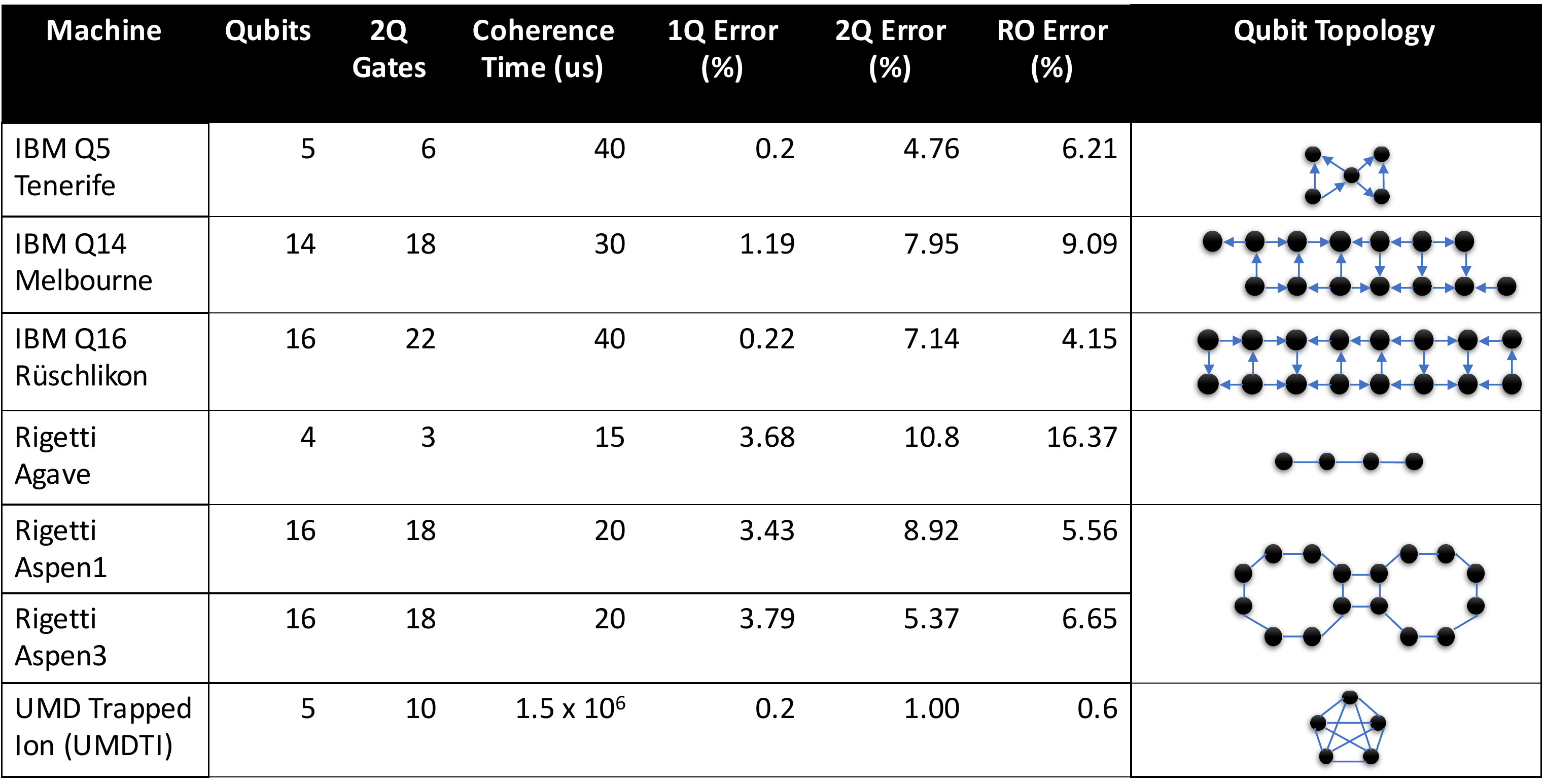}
     \caption{Characteristics of the devices used in our study. Each device has different qubit and gate count (higher is better), coherence time (higher is better), error rates (lower is better) and topology (dense connectivity is better). Rigetti Agave has 8 qubits in a ring topology, but only 4 qubits were available during our study.}
     \label{fig:machines}
 \end{figure*}
 \begin{figure*}[t]
    \centering
    \includegraphics[scale=0.42]{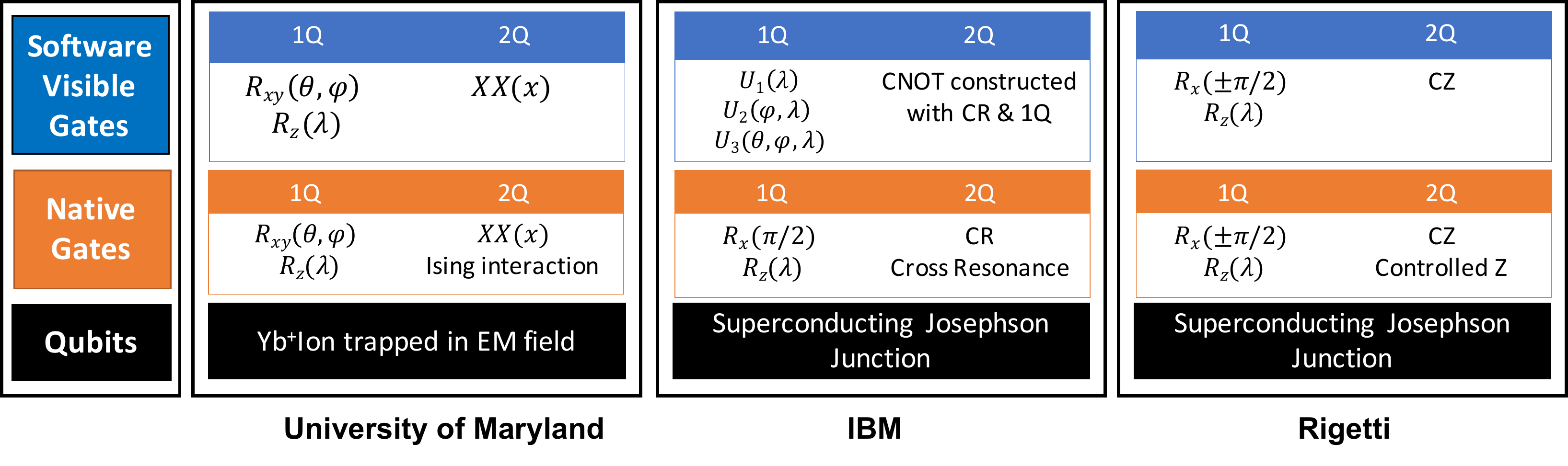}
    \caption{Qubit type, native gates and software-visible gates in the systems used in our study. Each technology lends itself to a set of native gates. Vendors may expose these gates in the software-visible interface or construct composite gates from multiple native gates.}
    \label{fig:qubit_prop}
\end{figure*}

As shown in Figure \ref{fig:machines}, the current candidate QC device technologies differ widely in key physical attributes.  First and foremost, the different methods of forming qubits are sufficiently distinct that even the fundamental gate operations performed on them differ widely.  (In contrast, consider that classical computers built from vacuum tubes, relay circuits, or transistors all hinge on a switch abstraction that maps similarly to Boolean logic gates.)  In addition to gate differences, there are also differences in how inter-qubit interactions are accomplished, and these lead to widely disparate communication approaches and connectivity topologies.  Finally, because of the differences in physical implementation, there are also considerable variations in the noise and error characteristics.  While all current QCs are susceptible to operation, communication, and measurement errors, the nature, magnitude, and spatiotemporal variance of these errors differs greatly from technology to technology.

This paper is the first to perform a cross-technology hardware software assessment of QC design. We assess how differences in fundamental gates, their software visibility, communication topologies, and noise characteristics all influence software performance and reliability on a range of real-system QC prototypes. To do so, we developed a full-stack cross-platform toolflow, \compilername, which allows us to start from QC programs written in a C-like high-level language \cite{scaffold, scaffcc1, scaffcc2}, progress through optimizations and mapping stages, and output device-specific code for {\em \syscnt different QC platforms from three different vendors employing two different underlying device technologies}.  We show how optimizing for specific device attributes can make significant improvements in performance and success rate. We further show how such device-specific characteristics can be input into an otherwise general/portable toolflow in order to allow multi-platform optimizations to be performed portably even with deeply device-specific optimizations. 

The contributions of this work include the following.  First, we perform the first multi-platform comparison of QCs from different vendors, built with different device technologies.  While it is far too soon to ``pick a winning technology'' from such comparisons, the purpose of our work is to build insights in how architecture and compiler design choices can best support the different technologies.  Conversely, we also believe that these early comparative studies offer important insights for device physicists as they roadmap further technology improvements \cite{cccqc, nasemreport}.
    
Second, to support our cross-platform studies, we have developed and will open-source the first multi-vendor QC compiler which compiles from high-level languages to multiple real-system QC prototypes, with device specific optimizations. \compilername takes as input a QC program written in a high-level language, a set of characteristics pertaining to implementation gates and communication topologies, and a summary of empirical device error data.  We show that although the device technologies vary, a common set of configuration parameters can express the gate set and noise data in a way that allows device-specific optimization through a portable, general toolflow.  We use this toolflow to compile 12 Scaffold benchmarks onto \syscnt real QC prototypes from 3 vendors employing two distinct qubit implementation approaches (superconducting and trapped ion). For the UMD ion trap system, \compilername is the first high-level language compiler.
    
Third, our evaluation offers architectural insights for future QC systems.  First and foremost, we demonstrate the importance of making software-visible the device technology's natural or fundamental gates. Shielding a technology's natural gates (e.g. the XX gate for trapped ion) by abstracting them into more familiar gates (e.g. CNOT) imposes runtime and error overheads that are too significant for current NISQ systems to easily overcome. Exposing device-specific gates also allows for additional compile-time optimizations of single-qubit operations.  Second, we find that even though trapped ion technologies have intrinsically lower error rates than many superconducting systems, there is still value in performing error-aware compilation for such systems.  
    
Fourth, our compilation approach melding device specificity with a common core toolflow offers very good results; we achieve portability without a tradeoff cost in performance or reliability.  In particular, \compilername outperforms vendor compilers.  On IBM devices with 5-16 qubits, \compilername provides geomean 3x (up to 28x) improvement in program success rate over the IBM Qiskit compiler \cite{qiskit}. On Rigetti devices with 4-16 qubits, \compilername provides geomean 1.45x (up to 2.3x) over the Rigetti Quil compiler \cite{quil}. \compilername obtains improvements over IBM and Rigetti compilers because of noise-adaptivity, optimizing qubit communication and optimizing single qubit operations. On the UMD ion trap computer (UMDTI), \compilername uses noise-adaptiveness to improve program success rates by up to $1.47$x, compared to a noise-unaware baseline. In particular, our paper is the first to demonstrate noise-adaptive compilation across 3 vendors and for trapped ion qubit technology. Finally, although compile time is not a primary design goal, \compilername scales well up to 72 qubits, the largest NISQ configuration announced thus far \cite{googlebristlecone}.

\section{QC Background}
\label{sec:background}

\subsection{Principles of Quantum Computing} 
A qubit is the fundamental unit of information in a QC system.
Qubits have two basis states $\ket{0}$ and $\ket{1}$, which are the analogues of the classical 0 and 1 states. However, quantum superposition allows a qubit to be in a complex linear combination, where its  state is $\alpha \ket{0} + \beta \ket{1}$, for $\alpha, \beta \in \mathbb{C}$. An $n$-bit QC system can potentially exist in a superposition state of $2^n$ basis states simultaneously, unlike classical registers which can be in exactly one of the $2^n$ values at any given time.
Qubits can be manipulated by modifying the complex numbers associated with the basis states, using operations which are commonly called gates. To obtain classical output, a qubit is measured, collapsing its state to either $\ket{0}$ or $\ket{1}$.

In a QC application, an algorithm is mapped to gates which execute on a set of qubits which are initialized appropriately. As the program executes, qubit amplitudes are manipulated and the state space is evolved towards the desired output. Finally, the qubits are measured or readout (RO) to generate classical output for the application.

\subsection{Quantum Gates}
Quantum gates are instructions which operate on one or more qubits. The functionality of a gate is achieved by applying some dynamic physical interaction (such as a microwave or laser pulse) to the qubit.  Complex quantum operations can be composed as a sequence of operations from a small set of universal gates. 
Universal QC systems, such as the ones we experiment on here, provide a universal set of single-qubit (1Q) and two-qubit (2Q) operations. 

The state of a single qubit can be represented by a complex vector on a unit sphere. All single-qubit operations can be viewed as rotation operations $R_x$($\theta$), $R_y$($\phi$) and $R_z$($\lambda$) along the X, Y or Z axes on this complex sphere. Rather than fully-general rotations, QC algorithms often use a  set of composite 1Q operations, such as X/NOT gate ($R_x$($\pi$)), Hadamard gate ($R_y$($\pi$/2)$R_z$($\pi$)) which generates superposition, Z gate ($R_z$($\pi$)) and others \cite{Mermin}.

2Q operations generate entanglement among qubits, resulting in non-classical correlated behaviour. Correlation from entanglement potentially allows a QC's state space to grow exponentially with qubit count. This is central to QC's power and is used by QC algorithms.
A common example of a 2Q gate is Controlled NOT (CNOT) which acts on a control and target qubit pair. 
 When the control qubit is in the state $\ket{1}$, the action of the gate is to flip the state of the target qubit\footnote{A CNOT gate with control C and target T is denoted as {\tt CNOT C, T}.}. Another example of a 2-qubit gate is a Controlled Z gate where the target qubit is rotated by $\pi$ radians along the Z axis if the control qubit is $\ket{1}$.
 

\subsection{NISQ Systems} 
Noisy Intermediate-Scale Quantum (NISQ) are near-term systems with less than 500-1000 qubits \cite{nisq}. This scale is typically too small to implement quantum error correction, but if used efficiently these machines may have promising applications in various domains \cite{nisq} and can pave the way towards practical QC. 

NISQ systems are built using a variety of qubit technologies, including superconducting qubits \cite{superc1}, trapped ions \cite{trappedion1}, spin qubits \cite{spin1}, among others. To reliably process information, these qubits should be ``coherent'' for sufficiently long, i.e., they should maintain the quantum state for a length of time. The qubits should also support sufficiently precise operations to allow the state to be manipulated correctly during the coherence window. To obtain useful output, qubits should also support accurate readout or measurement operations \cite{divincenzo}.  Because of the error rates in current NISQ systems, some fraction of the runs result in the wrong answer being calculated.  As a result, it is common to run a QC program many times with a figure of merit being the {\em success rate}, the fraction of runs that resulted in a correct answer.

\section{Device, Architecture Tradeoffs}
\label{sec:archquests}
\subsection{Native Gate Choices}
Figure \ref{fig:qubit_prop} shows the qubit technologies and gates (or operations) used in the systems at IBM \cite{ibmq_backends}, Rigetti \cite{rigetti_transmons} and UMD \cite{trappedion3}. IBM and Rigetti use superconducting qubits based on Josephson junctions, while UMD uses ions trapped in an electromagnetic field.  

In some ways, these different qubit implementation options are analogous to how classical computers might be implemented using vacuum tubes or CMOS transistors.  On the other hand, while many classical devices can all be abstracted as on-off switches, qubit technologies are still more distinct. Each QC vendor implements a set of \emph{native} operations that are feasible on their platform. Like classical NAND and NOR gates, all QC operations must be composable from a universal set of native gates. Typically, the vendor provides at least one 1Q and one 2Q operation. Figure \ref{fig:qubit_prop} shows that differences in underlying device technologies lend themselves to quite different native operations.  

In IBM Q systems, the fundamental 2Q interaction is a cross resonance gate where one qubit is driven at the resonant frequency of the other \cite{ibm_cr_gate}. On Rigetti systems, 2Q interactions are Controlled Z gates where a Z gate is applied on the target qubit when the control is $\ket{1}$. In the UMD system, entanglement is generated using Ising interaction (XX) which uses ion motion to couple the qubits.  

At a higher level, each vendor also decides on what sort of programmable machine interface to support.  This software-visible set of operations can include either native gates themselves or composite gates which use multiple fundamental gates.

Therefore, one key design decision is: {\em What sorts of operations should the machine expose to software?  How do these choices of software-visible operations affect the performance and reliability of the applications run on them?  And is there value in having these operations be unified across different implementations or is it more important to have them be well-tailored to underlying device characteristics?}
 
\subsection{Communication Characteristics}

As shown in Figure \ref{fig:machines}'s topology column, different QC implementations come with different qubit connectivity attributes.  For example, the IBM and Rigetti systems have sparse near-neighbor connectivity where 2Q operations can only be performed between {\em adjacent} qubits. To perform 2Q gates between non-adjacent qubits, these machines perform one or more SWAP operations to move relevant qubits until the control and target qubits for a 2Q operation are in adjacent locations. Each SWAP operation between two adjacent qubits (i.e., each hop) requires 3 2Q gates\footnote{For two qubits $A$ and $B$, {\tt SWAP(A,B) :=  \{CNOT A,B; CNOT B,A; CNOT A,B;\}}.}.  These SWAP operations are time-consuming, but even worse, each one is error-prone; a program with too many SWAPs is highly unlikely to get the right answer.  UMDTI natively supports full connectivity between any pair of ions, which means that swap operations are not required on this architecture.  

The distinction in communication topologies is not coincidental.  Indeed, full communication connectivity is easier to achieve in ion trap technologies at least at small qubit counts. 2Q gates in superconducting technologies require a physical resonator immediately between the 2 qubits to perform the operation, whereas in ion trap approaches, the interacting ions need not be physically adjacent and do not need physical resonators between them when the laser-pulsed operation occurs.  

{\em Our work is the first to perform comprehensive real-systems measurements between high-level language applications running on superconducting and ion trap implementations.  We ask and answer questions regarding how different communication topologies affect application success rate and runtime.  We also explore the best methods for a common toolflow to target such widely-divergent technologies and topologies. }

\subsection{Noise and Coherence Characteristics}

\begin{figure}
    \centering
    \includegraphics[scale=0.5]{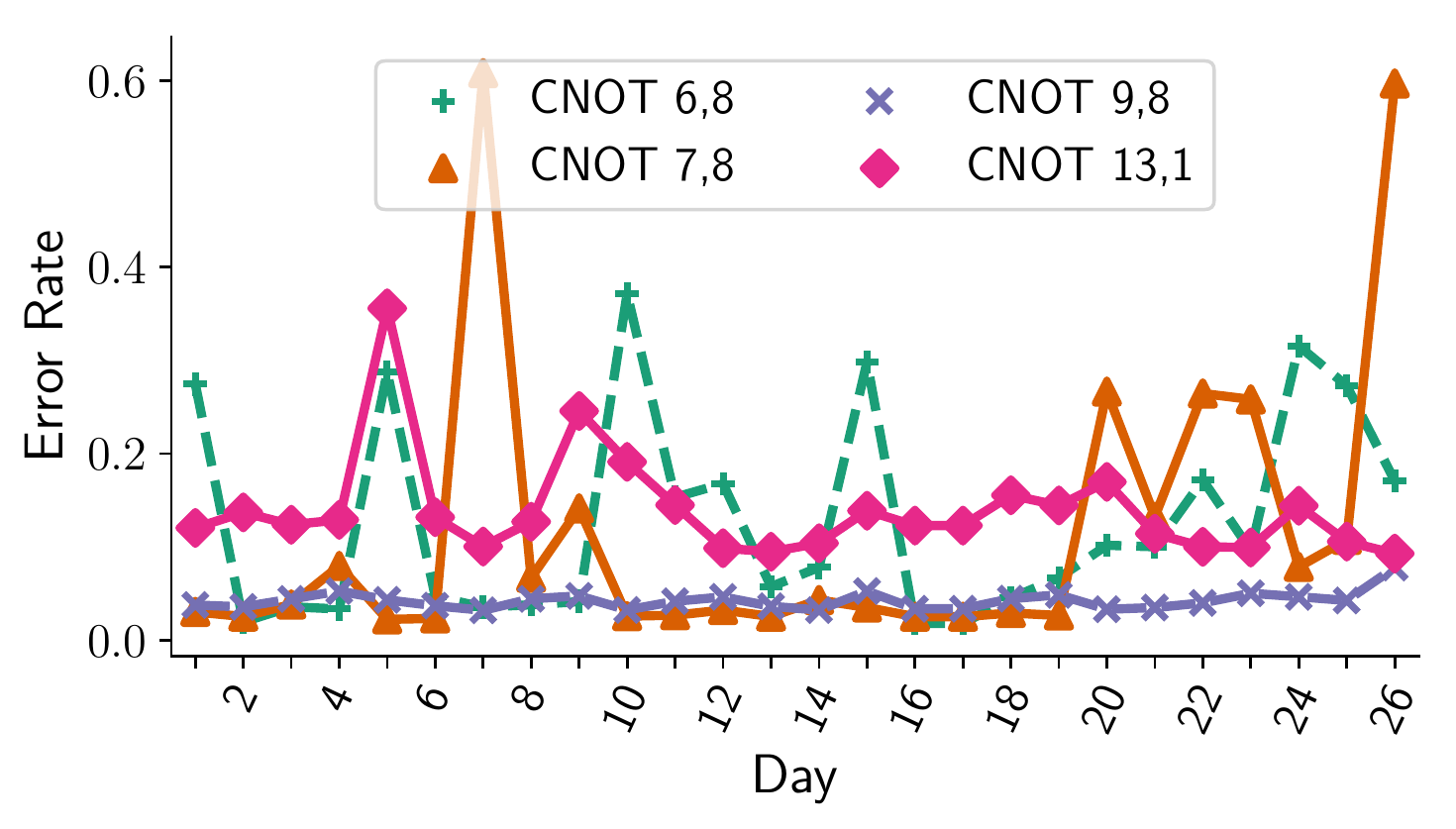}
    \caption{Daily variation of error rates of 4 hardware supported 2Q (CNOT) gates in IBMQ14. 2Q error rate in this system averages 7.95\%, but varies 9x across qubits and days.}
    \label{fig:ibm_cx_var}
\end{figure}
Finally, QC systems today are characterized by high and often fluctuating error rates (or ``noise'') in their gate operations and qubit readouts \cite{asplos, superconducting_stability}. NISQ systems have imprecise operations with high 1Q, 2Q and readout (RO) errors. 1Q error rates are smaller than 2Q and RO, but present. 2Q operations are fundamentally harder to perfect and are often composed to multiple 1Q operations. The superconducting qubits used in IBM and Rigetti have defects arising from their lithographic manufacturing processes and their variance is only beginning to be understood \cite{superconducting_stability}. The IBM qubits are calibrated twice a day and the experimental measurements of 1Q, 2Q and RO error rates are posted online \cite{ibmqexp}, as Figure \ref{fig:ibm_cx_var} shows for \ibmnameshort. Across IBM and Rigetti systems, 2Q and RO error rates vary up to 9x across qubits and calibration cycles. Ion traps have less temporal variance, and UMDTI has error rates lower than the machines based on superconducting qubits.  Nonetheless, despite lower error rates, there is still spatial and temporal variance of 1-3\% in 2Q error rates across the qubits in the system \cite{trappedion3}. These differences arise from the difficulty in qubit control using lasers and their sensitivity to motional mode drifts from temperature fluctuations. 

Another figure of merit in QC systems is the {\em coherence time}, which is a fundamental limit on the time up to which information can be reliably manipulated on a qubit. A loose analogy might be made to the DRAM refresh interval in classical systems.  Figure \ref{fig:machines} shows that IBM and Rigetti systems have short coherence time compared to UMDTI. Nonetheless, as QC systems are able to run longer and larger programs successfully, the limits imposed by finite coherence time will play a role in how algorithms are developed and how programs are compiled.

Therefore, we ask: {\em Can a single toolflow be built that maps well across widely varying implementations, and that optimizes algorithm mappings to the underlying implementation's error characteristics and coherence intervals?  Moreover, are there a small, common set of figures of merit that can guide optimization across such divergent implementations?}  

\subsection{Our Work}

Based on the prototype device technologies and their architectural implications, this work first develops a common toolflow, TriQ, that maps well to widely-different  machine implementations.  Using TriQ, we ask and answer questions about gate sets, communication tradeoffs, and other architectural choices on these machines.  We quantify our answers by presenting {\em real-system} runs on \syscnt different QC implementations based on two distinct implementation technologies.

\section{Design and Overview of \compilername}
Many traditional compilers are structured as a language-dependent front-end, a hardware- or ISA-dependent back-end, and a set of neutral analysis passes in the middle (intermediate representation or IR). The abstractive power of such approaches shields the higher-level optimizations from many hardware-specific details and vice versa. In contrast, our work here requires that many more hardware and software implementation attributes are available to all or nearly all of the full-stack of the compiler.

\begin{figure*}
\centering
\begin{minipage}{.58\textwidth}
  \centering
      \includegraphics[scale=0.4]{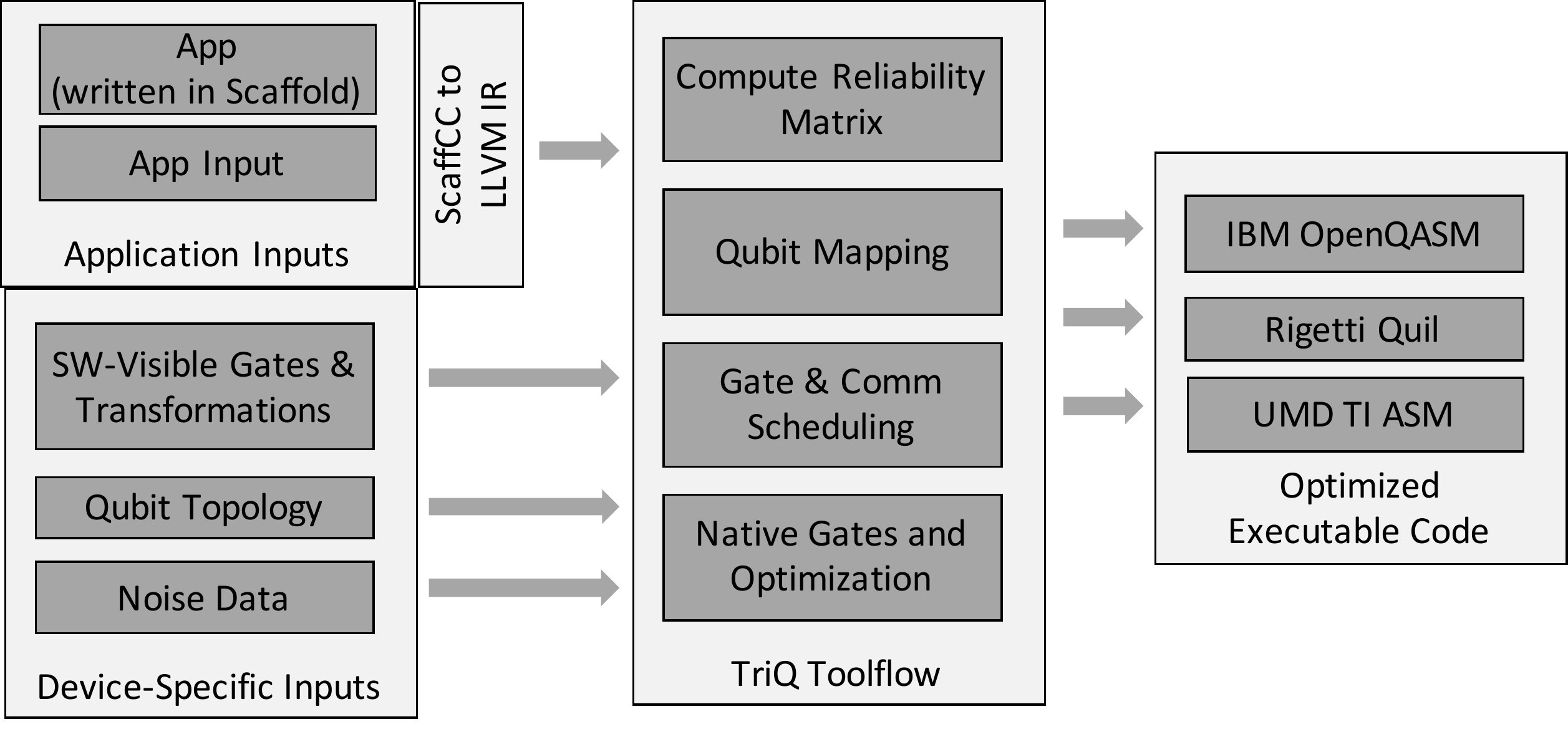}
    \caption{Overview of the \compilername toolflow. Input to \compilername consists of high-level Scaffold programs and their inputs, as well as device-specific QC system properties.  Output is optimized code in one of three vendor-specific executable formats.  
    }
    \label{fig:toolflow}
\end{minipage}%
\hfill
\begin{minipage}{.38\textwidth}
  \centering
      \includegraphics[scale=0.4]{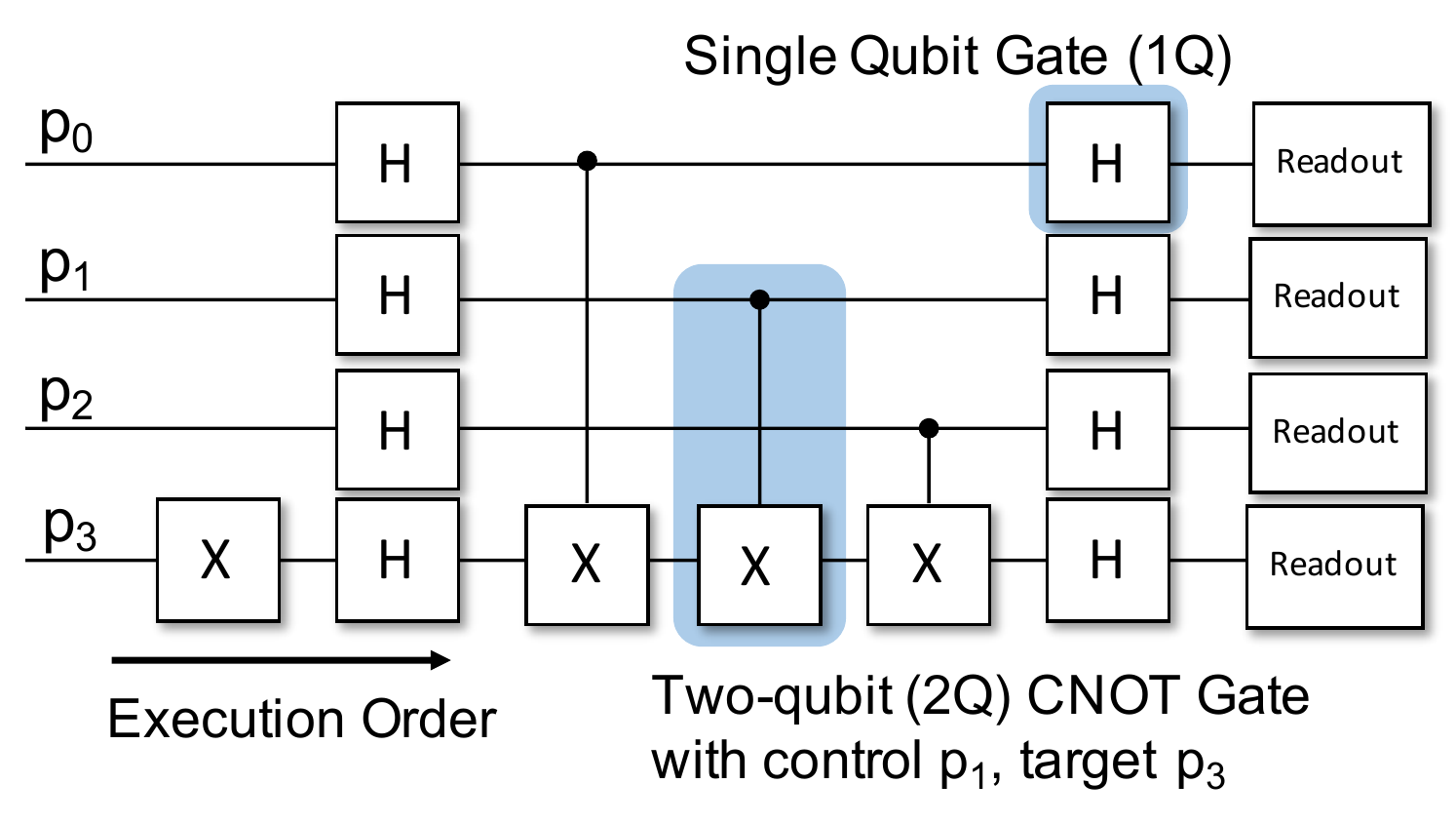}
    \caption{IR-level circuit diagram for Bernstein-Vazirani with 4 qubits (BV4). The program IR depicts program qubits and has 1Q, 2Q and RO operations.}
    \label{fig:bv_ir}
\end{minipage}
\end{figure*}

\subsection{Overview}
As shown in Figure \ref{fig:toolflow}, the \compilername toolflow contains core functionality to perform noise-aware qubit mapping, 1Q optimizations, and communication optimizations.  For all this functionality, the device-specific attributes are provided to the core functionality as compiler inputs.  This includes the machine's qubit count and connectivity, its native gate set, and a summary of its noise characteristics.  In essence, it operates like a multi-target compiler in which characteristics such as the ISA and operation latencies are provided as compile-time inputs. In this way, our compiler can target very different devices simply by changing input characteristics.  Section \ref{sec:results} shows that this flexibility comes with no performance trade-off; \compilername outperforms the native vendor compilers in both performance and success rate. 

Our toolflow accepts program inputs written in Scaffold, a C-like language that represents QC programs in a device- and vendor-independent manner. The ScaffCC compiler \cite{scaffcc1, scaffcc2} parses from Scaffold into an LLVM IR \cite{llvm}  consisting of 1Q and 2Q gates. Figure \ref{fig:bv_ir} shows the IR for the Bernstein-Vazirani algorithm \cite{bernsteinvazirani}, a common NISQ benchmark program. ScaffCC automatically decomposes higher-level QC operations such as Toffoli gates into native 1Q and 2Q representations. Since QC programs are usually compiled for a fixed input, ScaffCC also takes the application input data and resolves all classical control dependencies.  The output IR graph includes the qubits required for each operation and data dependencies between operations.  



From both the application inputs (program and input data) and the inputs about the device characteristics (resource counts, noise statistics, etc.) the core compiler passes analyze and optimize mappings before generating device-specific executable code to be run on one of \syscnt real systems.  The sequence of compiler analyses is discussed below. In our experiments, we vary which optimizations are applied. Table \ref{tab:configurations} names and summarizes the optimization approaches.

\begin{table}
\caption{Compilers and optimization levels considered.}
\label{tab:configurations}
\small
\begin{tabular}{|l|l|}
\hline
Compiler                 & Description                                                                                                              \\ \hline \hline
\noopt    & \begin{tabular}[c]{@{}l@{}} \compilername. No optimization. Default qubit mapping \end{tabular}  \\ \hline
\oneqopt  &
\begin{tabular}[c]{@{}l@{}} \compilername, 1Q gate optimization. Default qubit mapping \end{tabular} \\ \hline
\commopt  & \begin{tabular}[c]{@{}l@{}} \compilername.  1Q opt. Communication-optimized mapping, \end{tabular}                  \\ \hline
\noiseopt & \begin{tabular}[c]{@{}l@{}} \compilername. 1Q opt. Comm- and Noise-optimized mapping.\end{tabular}              \\ \hline
\Qiskit   & IBM Qiskit compiler version 0.6.0 \cite{qiskit} \\ \hline
\Quil     & Rigetti Quil compiler version 1.9  \cite{pyquil}   \\
\hline
\end{tabular}
\normalsize
\end{table}

\subsection{Reliability Matrix Computation}
\label{sec:qubit_reliablity_computation}
The qubit mapping and communication orchestration phases must determine good spatial placements for qubits and good routing paths for 2Q gates.  As Figure \ref{fig:machines} shows, 2Q and RO operations dominate error rates and are important to optimize for \cite{asplos}.  The gate errors on both superconducting and trapped ion prevent long gate sequences and are more limiting than coherence times. For these reasons, a central aspect of qubit mapping and gate orchestration decisions is optimizing for the reliability of 2Q and RO operations. The challenge is doing so in a way that ports well across very different implementations.

\begin{figure}[t]
    \centering
    \subfloat[Example 8-qubit device]
    {
    \includegraphics[scale=0.25]{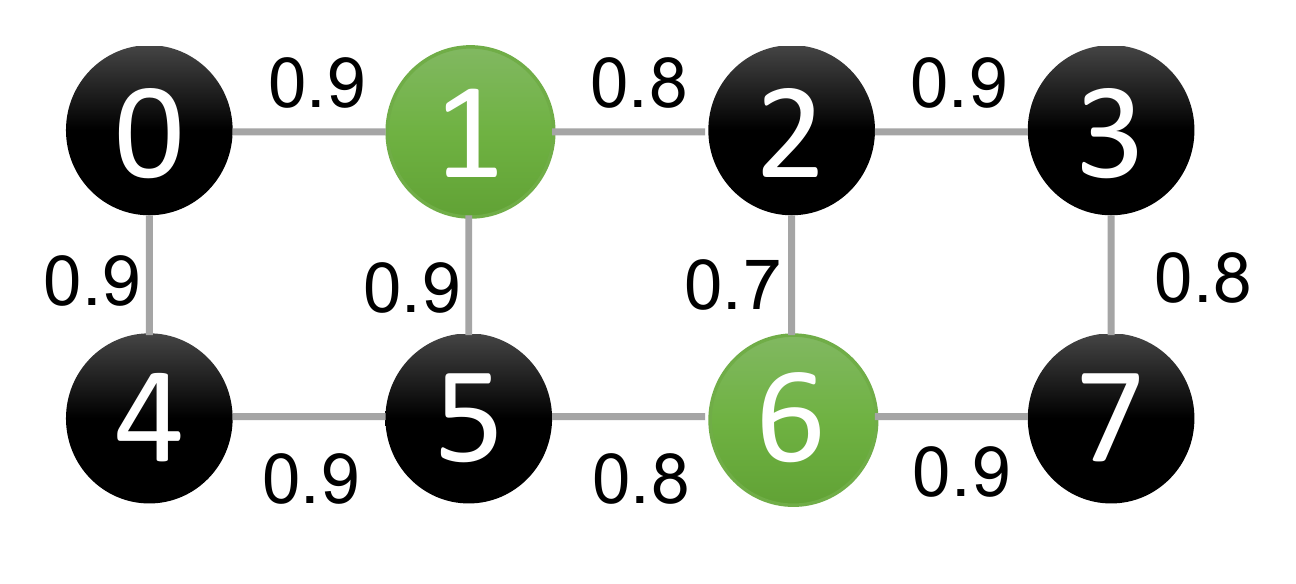}
    \label{fig:rmatrix_8q_device}
    }
    \subfloat[2Q reliability matrix]
    {
    \includegraphics[scale=0.3]{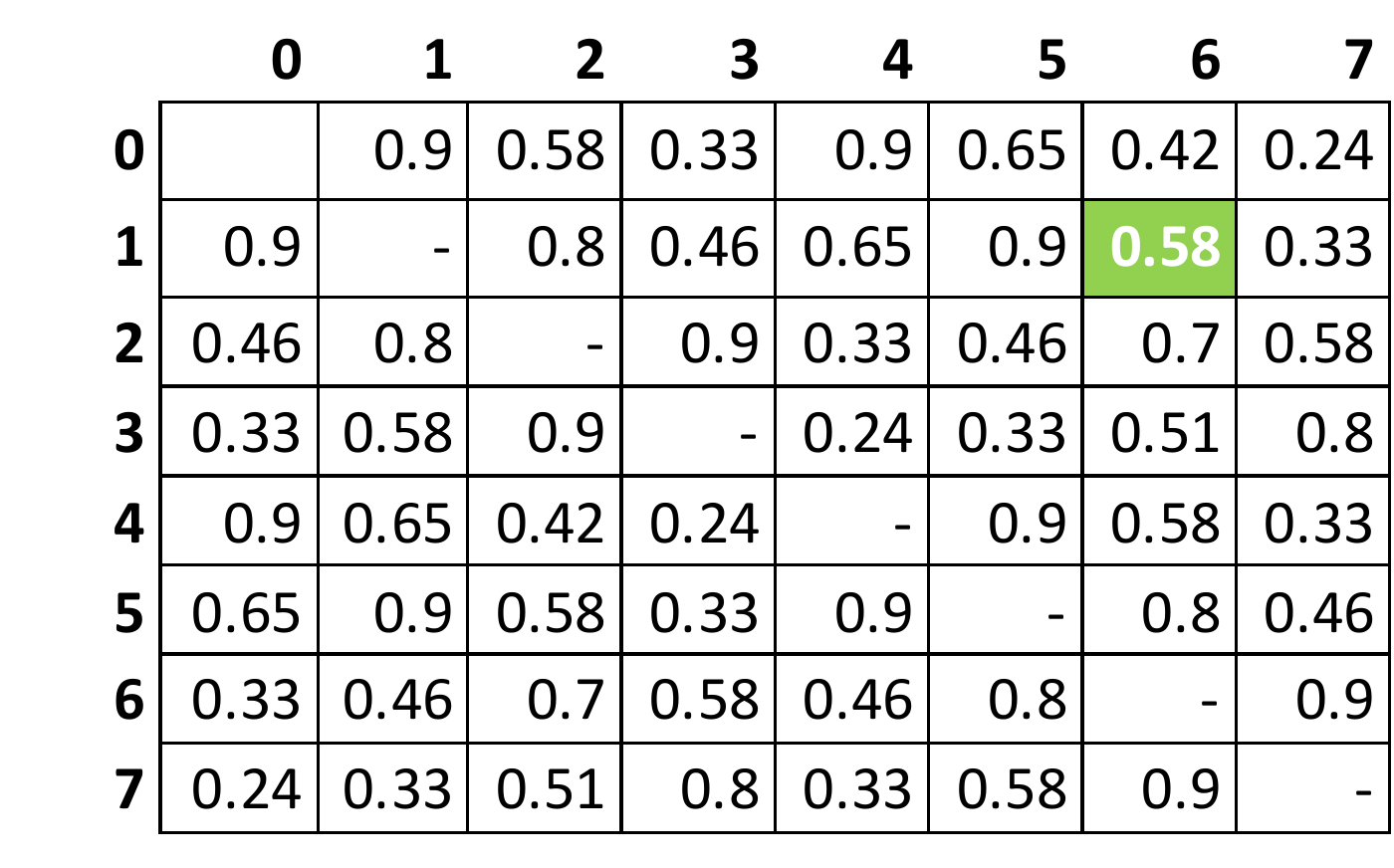}
    \label{fig:rmatrix_table}
    }    
    \caption{Example 8-qubit device with 2Q gate reliabilities and corresponding 2Q reliability matrix. For a 2Q gate involving qubits 1 and 6, the highest reliability path involves first swapping 1 and 5 (reliability = $0.9^3$) and then performing the 2Q gate. Thus, the (1,6) entry of the reliability matrix is $0.9^3*0.8$ or 0.58.}
    \label{fig:reliabmatrix}
\end{figure}

To inform qubit mapping and communication orchestation, \compilername uses the provided qubit topology and noise data to construct a  matrix which summarizes the ``end-to-end'' reliability of 2Q operations between any pair of qubits, including any communication routing required to co-locate the qubits. Figure \ref{fig:reliabmatrix} shows an example. The $(i,j)^\text{th}$ entry in this matrix estimates the reliability of performing a 2Q operation from qubit $i$ to qubit $j$, including the cost of communication. By distilling the important factors of a machine's topology and 2Q errors into single matrix representation, this approach is applicable both to fully-connected machines like UMDTI as well as to machines with more limited topologies. 

To fill in the reliability matrix, \compilername considers the topology of the machine, where nodes are hardware qubits and edges are hardware-supported, direct 2Q gates between them. Each edge is labelled with the reliability of the corresponding 2Q gate. To estimate the reliability score for non-local operations, \compilername performs an all-pairs swap cost computation using the Floyd-Warshall algorithm \cite{clrs}. For each pair of qubits $c$ and $t$, it determines the most reliable neighbor $t'$ of $t$. $t'$ is the neighbor which maximizes the product of reliability of the swap path from $c$ to $t'$ and the 2Q gate from $t'$ to $t$ (for IBM machines, we also include the error rates of any extra 1Q gates necessary for orienting the gates in the hardware-supported directions). In Figure \ref{fig:reliabmatrix}, the best neighbor for a 2Q gate from 1 to 3 is 4. 




For noise-unaware compilation such as \commopt, the 2Q gate reliability for all edges is set as the average error rate in the system. Then the reliability matrix computation in effect determines shortest paths which minimizes hop count. The noise-aware \noiseopt approach uses the input noise/error data to set the gate reliability. Here, the shortest path computation choose the most reliable path, minimizing prevalence (severity and count) of erroneous operations.For example, if the two program qubits $p_0$ and $p_3$ from Figure \ref{fig:bv_ir} are mapped to 
qubits $1$ and $3$ in Figure \ref{fig:reliabmatrix}, the reliability of a 2Q operation between them is 0.58.  We also record the readout reliabilities for the hardware qubits in a vector, where the $i^\text{th}$ entry denotes the accuracy of measuring the state of qubit $i$.  This matrix becomes an input for the subsequent passes that follow.

\subsection{Qubit Mapping}
\label{sec:qubit_mapping}
To map program qubits to hardware qubits, \compilername uses a constrained optimization method similar to \cite{asplos}. The optimization creates variables for each program qubit denoting which hardware qubit it may be mapped to. The reliability matrices for 2Q and readout operations previously discussed ascribe possible operation costs for each program operation. The optimization goal is to maximize success rate, so our objective function is a function of the reliability for the program dependence graph as mapped.

Maximizing this objective implies that communicating qubits should be mapped close together, and hardware gates and readout units which have poor reliability scores should be avoided. The optimization problem can be solved by expressing the variables, constraints and objective in a Satisfiability Modulo Theory solver \cite{z3, omt_z3}. Such approaches have been used to find optimal hardware mappings for classical programs \cite{Nowatzki:2013:GCS:2491956.2462163} and for the IBMQ16 device by \cite{asplos}.

Our mapper is more general and more scalable than prior QC work. First, \cite{asplos} uses a problem formulation which assumes that the device is a 2D grid. Since \compilername is multi-platform, it must target devices with arbitrary topology, which our 2D reliability matrix supports well. In addition, our objective function is chosen to be more scalable than prior work, while still targeting reliability.  In particular, our objective function {\em maximizes the minimum reliability of any operation in the mapped graph}. In contrast, prior QC work has used a reliability product across the whole graph; these require more exhaustive search techniques that slow down the SMT solver runs. (We use the product-style reliability estimate in populating the reliability matrix, but use the faster maximize-the-minimum approach for the SMT solver that maps based on the matrix.) Our implementation allows the SMT solver to prune bad solutions early in the search tree: if it maps two qubits and a gate has lower reliability than the current optimum, the mapping can be discarded. In contrast, for the product-based method in \cite{asplos}, the solver must place all qubits and evaluate the product over some or all operations before it can discard a mapping as sub-optimal. As a result, \compilername is more scalable, and offers acceptable success rates on par with prior work.  While different objective functions can be employed as desired, the current \compilername approach seems particularly useful for problems where \cite{asplos} fails to compute a solution. Using the routing method in the next section, \compilername also avoids unnecessary swaps incurred by \cite{asplos} and improves the likelihood of program success up to $50\%$ on IBMQ16.
\subsection{Gate and Communication Scheduling}
We schedule gates in a topologically-sorted order from the IR's gate dependencies. This ensures that before a gate is executed, all its dependent gates are completed. For example, in Figure \ref{fig:bv_ir}, the X gate is executed first on $p_3$, followed by H gates which can execute in parallel on all qubits, followed by the other gates. Once scheduled, where selected qubits are non-adjacent, the compiler determines routing paths for 2Q gates. (This step is not required for fully-connected topologies such as UMDTI.) The compiler seeks to use the most reliable path for the control and target qubit pair based on Section \ref{sec:qubit_reliablity_computation}'s reliability matrix. The compiler inserts SWAP gates to move the qubits along the best path, culminating in the desired local 2Q operation. \compilername updates the qubit mapping to reflect the swaps and processes the next 2Q gate using the new mapping. 
\subsection{Gate Implementation, Optimization and Code Generation}
To generate executable code, \compilername must translate from higher-level IR gates into the software-visible gates of the device. Using the legal transformations provided as input to the toolflow, the compiler replaces SWAPs, CNOTs, and other operations in terms of the native or software-visible gate set. For example, on superconducting machines, to implement a {\tt SWAP A,B} gate, it is decomposed into 3 2Q gates as {\tt CNOT A,B; CNOT B,A; CNOT A,B}. 

For IBM, CNOT is a machine-supported, software-visible operation and needs no further transformation, although access to lower-level native gates might allow for further optimizations. For Rigetti, CNOT is not software visible and instead, we decompose a {\tt CNOT A,B} gate into a sequence of rotations and CZ operations: {\tt $R_z$($\pi$/2) B; $R_x$($\pi$/2) B; $R_z$($\pi$/2) B; CZ A,B; $R_z$($\pi$/2) B; $R_x$($\pi$/2) B; $R_z$($\pi$/2) B;}. Similarly, for UMDTI, we decompose the CNOT as 
{\tt $R_y$($\pi$/2) A; $XX$($\pi$/4) A,B; $R_y$($-\pi$/2) A; $R_x$($-\pi$/2) A; $R_z$($-\pi$/2) A;}.

\compilername also applies device-specific transformations at this stage. For IBM devices with directed CNOTs, \compilername orients the IR CNOTs in correct direction using additional 1Q operations \cite{Mermin}. 
Next, \compilername optimizes sequences of 1Q gates. For each qubit, \compilername finds continuous sequences of 1Q operations. Since 1Q operations are rotations, each 1Q gate in the IR can be expressed using a unit rotation quaternion which is a canonical representation using a 4D complex number. \compilername composes rotation operations by multiplying the corresponding quaternions and creates a single arbitrary rotation. This rotation is expressed in terms of the input gate set. Furthermore, on all three vendors, Z-axis rotations are special operations that are implemented in classical hardware and are therefore error-free. 
\compilername expresses the multiplied quaternion as a series of two Z-axis rotations and one rotation along either X or Y axis \cite{quaternion, qiskit}, thereby maximizing the number of error-free operations.

\subsection{Executable Generation}
Following the stages in order as described above, the toolflow generates code in a format that is executable on the targeted real-system platform.  For IBM, this is OpenQASM \cite{openqasm1}, for Rigetti it is Quil \cite{quil}, and for UMDTI, there is a special low-level assembly code syntax we target.  We stress that the key analysis and optimization functionality is all in the core toolflow.  The device-specific code generation backend is merely intended to output executable code in a syntax supported by that machine.
\section{Experimental Setup}
\begin{figure}
\centering
\includegraphics[scale=0.3]{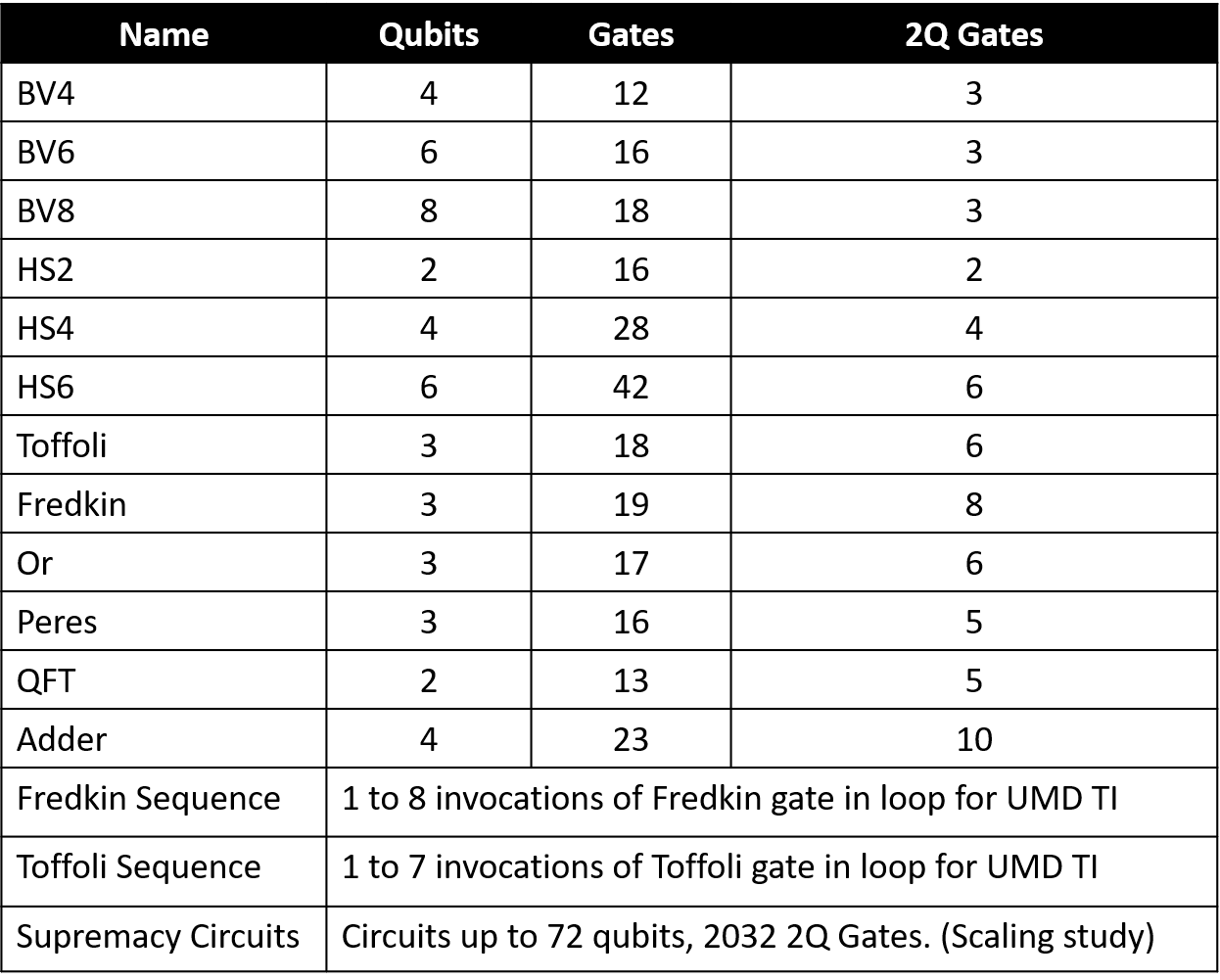}
\caption{Summary of benchmarks used in our study. The supremacy circuits are used only for scaling studies.}
\label{fig:benchmarks}
\end{figure}

{\bf Benchmarks:} Figure \ref{fig:benchmarks} lists the QC programs in our study. Used in prior work on NISQ system evaluation and compilation \cite{asplos, bench1, bench2, bench3}, these benchmarks include the Bernstein-Vazirani algorithm \cite{bernsteinvazirani}, Hidden Shift Algorithm \cite{hiddenshift}, Quantum Fourier Transform \cite{NielsenChuang}, an adder and multi-qubit QC gates such as Toffoli and Fredkin \cite{Mermin}. We use these benchmarks because they constitute an essential part of many large QC applications \cite{shor1, grover}. We created Scaffold programs for each benchmark and obtained LLVM IR using the ScaffCC compiler \cite{scaffcc1}. UMDTI's lower error rates support longer gate sequences, so we also created longer benchmarks by running the Toffoli and Fredkin benchmarks in a loop, similar to patterns in applications such as Grover's search \cite{grover}. 

To study scalability, we used supremacy circuits from Google's Cirq tool \cite{boxio, cirq}. These circuits are large programs designed to demonstrate the computational capability of QC systems. We experiment with supremacy circuits with up to 72 qubits and 2032 two-qubit gates (depth 128). 

{\bf Compilers:}
Table \ref{tab:configurations} summarizes the compilers used in our study and their optimization levels.
\noopt generates code in terms of a technology-independent gate set (CNOTs and common 1Q gates) and naively translates it to the software-visible gates on the device. 
\oneqopt generates code in the software-visible gate set of the machine and optimizes it using the properties of the gate set. \commopt uses a communication optimized mapping using a reliability matrix constructed from ideal average error rates. In \noiseopt, the mapping is optimized using both noise and topology data by using a reliability matrix with gate errors from calibration data. For comparison, we use IBM's Qiskit compiler/mapper version 0.6.0 and Rigetti's Quil version 1.9 as baselines. These were the latest compiler versions while performing the experiments.  UMDTI does not have a high-level compiler. Our compilation experiments use an Intel Skylake processor (2.4GHz, 128GB RAM) using gcc version 5.4 and Python3.5. Our toolflow uses the C++ APIs of Z3 SMT solver version 4.8.3 \cite{z3} for implementing the qubit mapping phase. 

{\bf Real-System QC Experiments:}
As already discussed in Figure \ref{fig:machines}, we performed empirical experiments on \syscnt operational QC machines from three organizations: IBM, Rigetti, and UMD.  The IBM experiments were performed by running on the IBM Quantum Experience \cite{ibmq, ibmqexp}. The IBMQ APIs also provide access to the daily machine calibration data, including error rates such as 1Q, 2Q and RO errors.  For the experiments on Rigetti and UMDTI, we worked with the operators of those machines to launch experimental runs, and we obtained similar calibration data directly from each vendor. Experiments on Rigetti were performed using the pyQuil APIs \cite{pyquil}. Compiled executables for UMDTI were prepared by us using gate calibration data from UMD. The system staff at UMD ran the compiled code on the device without modifications.

QC machine time and availability is scarce and variable across vendors. In some experiments, we were limited to testing selected compiler configurations. In addition, the Rigetti Agave, UMDTI and IBMQ5 machines did not have enough qubits at the time of our experiments to accommodate the BV6, BV8 and HS6 benchmarks. 

Before each experiment, we recompile the benchmarks using the latest calibration data. For IBM and Rigetti machines, we use 8192 trials for each benchmark run. For UMDTI, we used 5000 trials per run because of less noise variability on this system. Success rate refers to the fraction of those repeated trials which give the correct answer. For example, success rate of $0.9$ means that $90\%$ of the trials produced the correct answer. For success rates, results from different graphs may vary because of experimental conditions. Results within a single graph are performed closely in time and so are comparable.
\section{Results}
The \compilername toolflow allows us to study opportunities for multi-platform optimizations, as well as architectural implications related to different device and implementation choices.  This section offers empirical real-system results on initial key questions.

\label{sec:results}
\begin{figure*}[th!]
    \centering
    \subfloat[IBMQ14]
    {
    \includegraphics[scale=0.25]{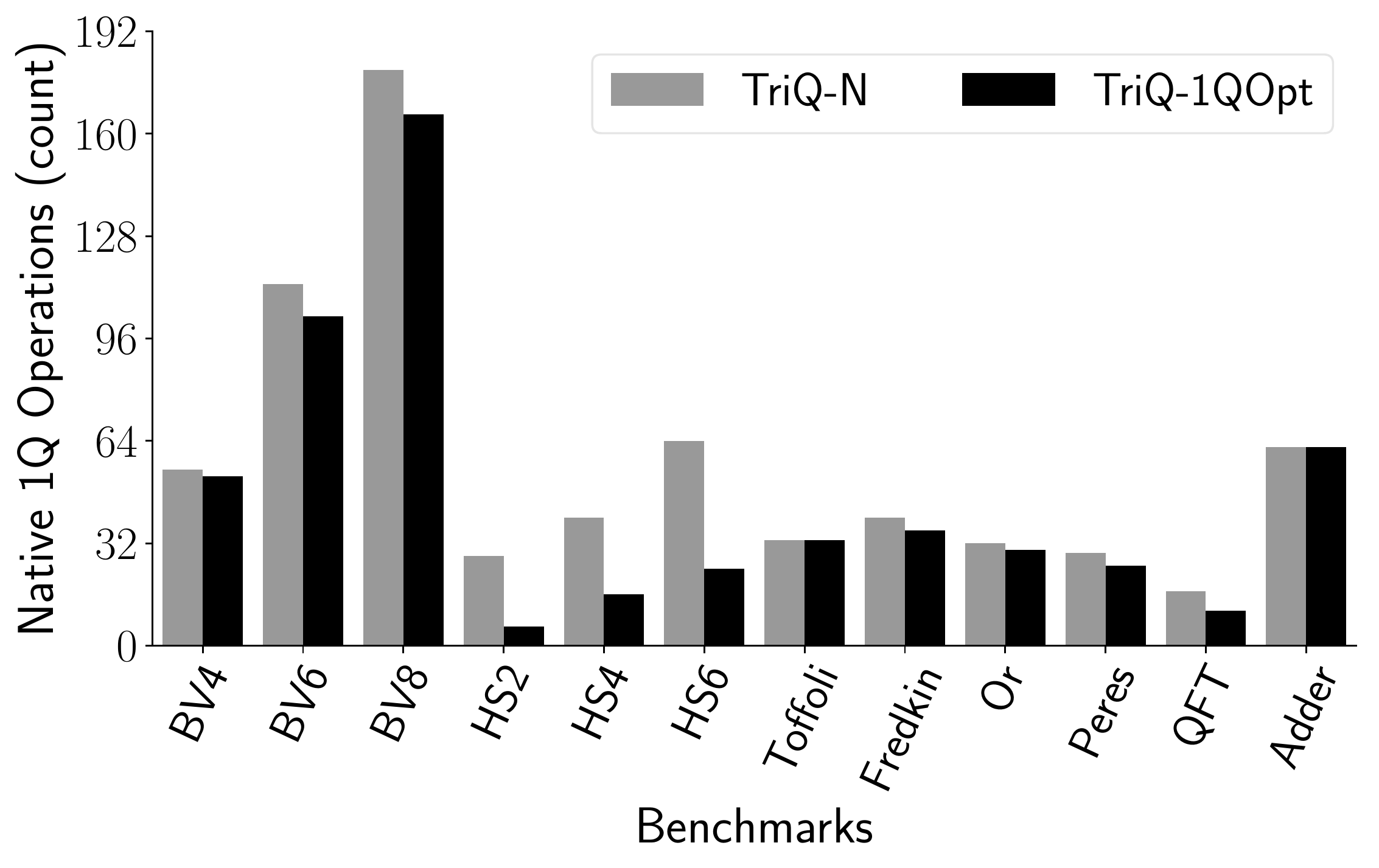}
    \label{fig:e1_gs_ibmq14_cnt}
    }
    \subfloat[Rigetti Agave]
    {
    \includegraphics[scale=0.25]{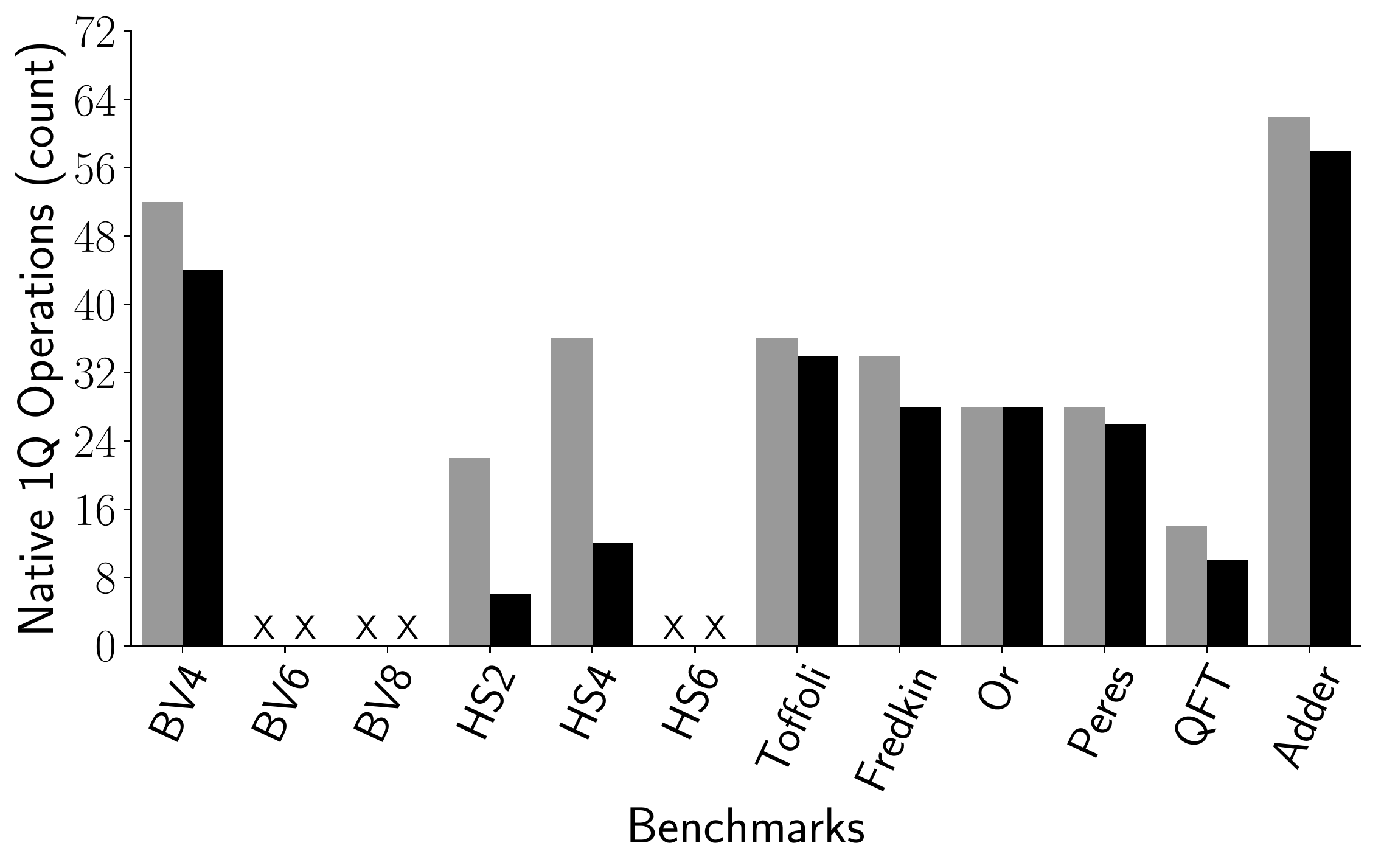}
    \label{fig:e1_gs_rigetti_cnt}
    }
    \subfloat[UMDTI]
    {
    \includegraphics[scale=0.25]{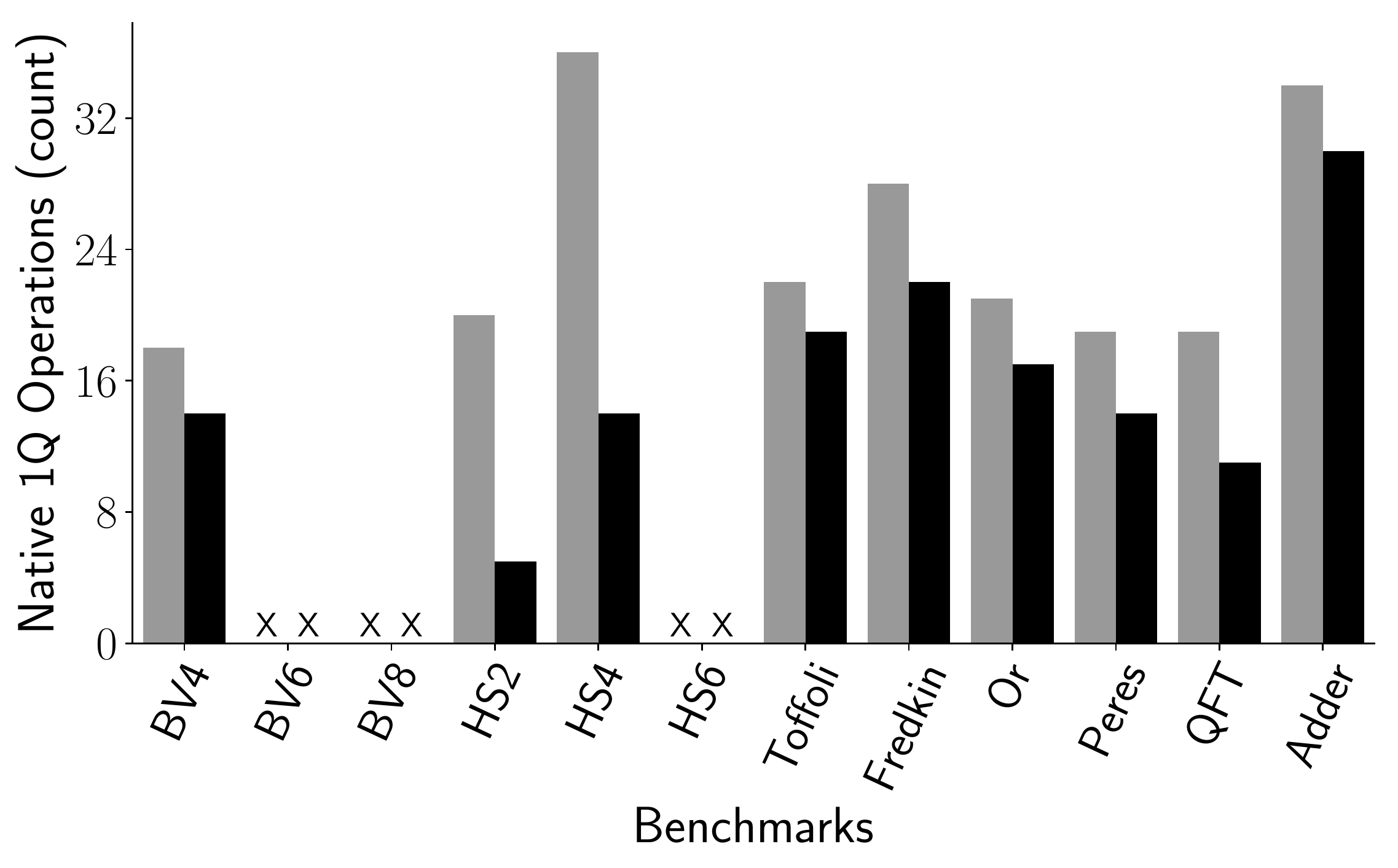}
    \label{fig:e1_gs_tion_cnt}
    }
    
    \caption{\textbf{Importance of single qubit gate optimization.} Counts of native 1Q operations (actual X and Y pulses applied on the qubits) when the compiler optimizes gates using properties of the gate set. \oneqopt reduces operation count by up to 4.6x by decomposing and optimizing composite gates. In (b) and (c), HS6, BV6 and BV8 are marked ``X'' because of size restrictions on Agave and UMDTI.}
    \label{fig:e1_gs_cnt}
\end{figure*}
\subsection{Gate Specificity and Optimizations}

Figure \ref{fig:e1_gs_cnt} shows the number of native 1Q operations using \noopt and \oneqopt.  Because so-called ``virtual Z gates" can be applied on all three vendors using runtime classical transformations, those rotations are error-free on all 3 vendors; we plot X and Y here. \noopt produces output code in terms of the software-visible gates, it does not perform any optimization. The native gate set plays a key role in the number of 1Q operations required by the unoptimized code. Namely, where swaps end up getting translated to sequences including native 1Q operations, then some benchmarks such as BV8 that require long swap paths on IBM will have a large number of supporting 1Q gates.

\begin{figure}[t]
    \centering
    \subfloat[IBMQ14]
    {
    \includegraphics[scale=0.25]{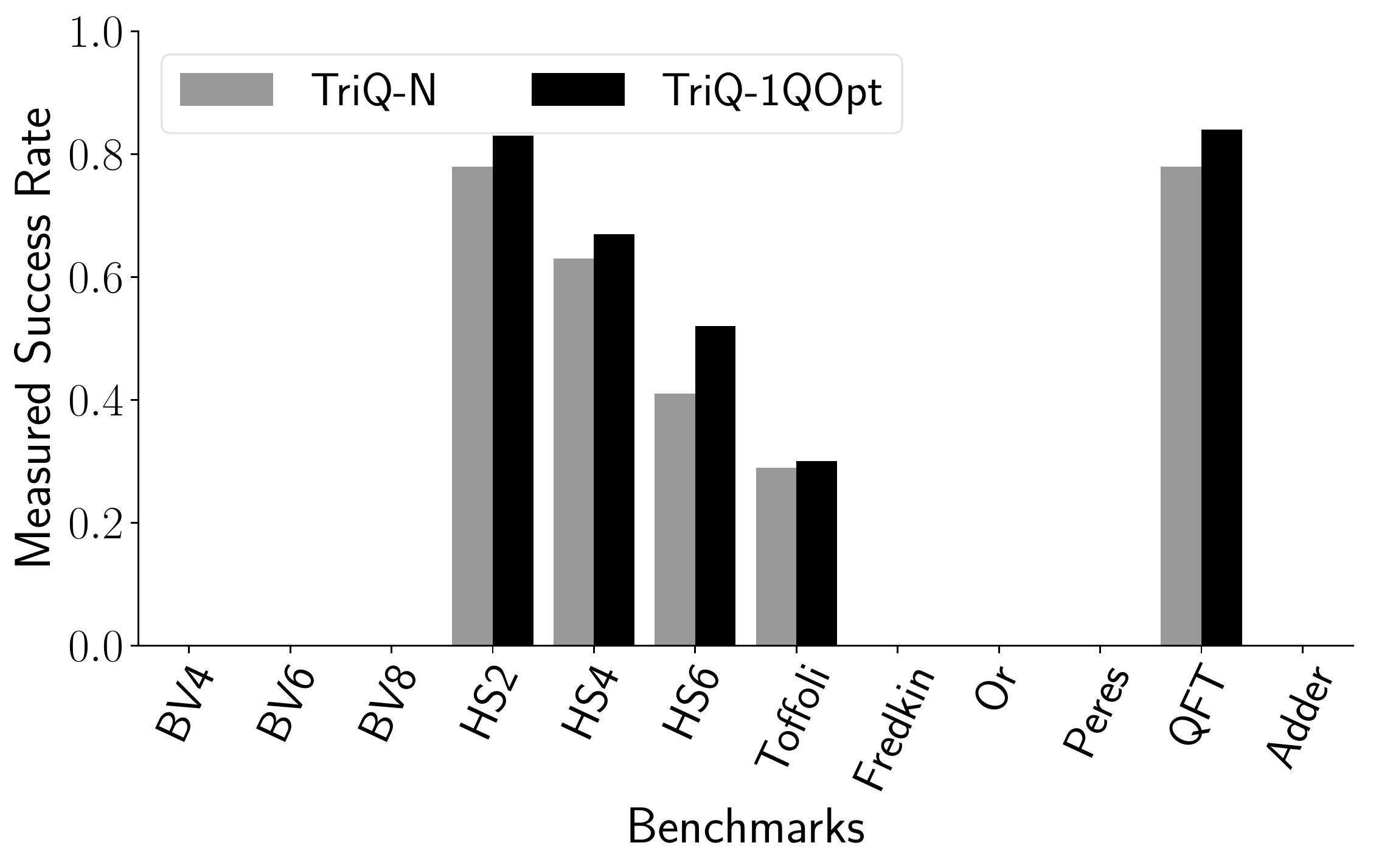}
    \label{fig:e1_gs_ibmq14_succ}
    }
    
    \subfloat[UMDTI]
    {
    \includegraphics[scale=0.25]{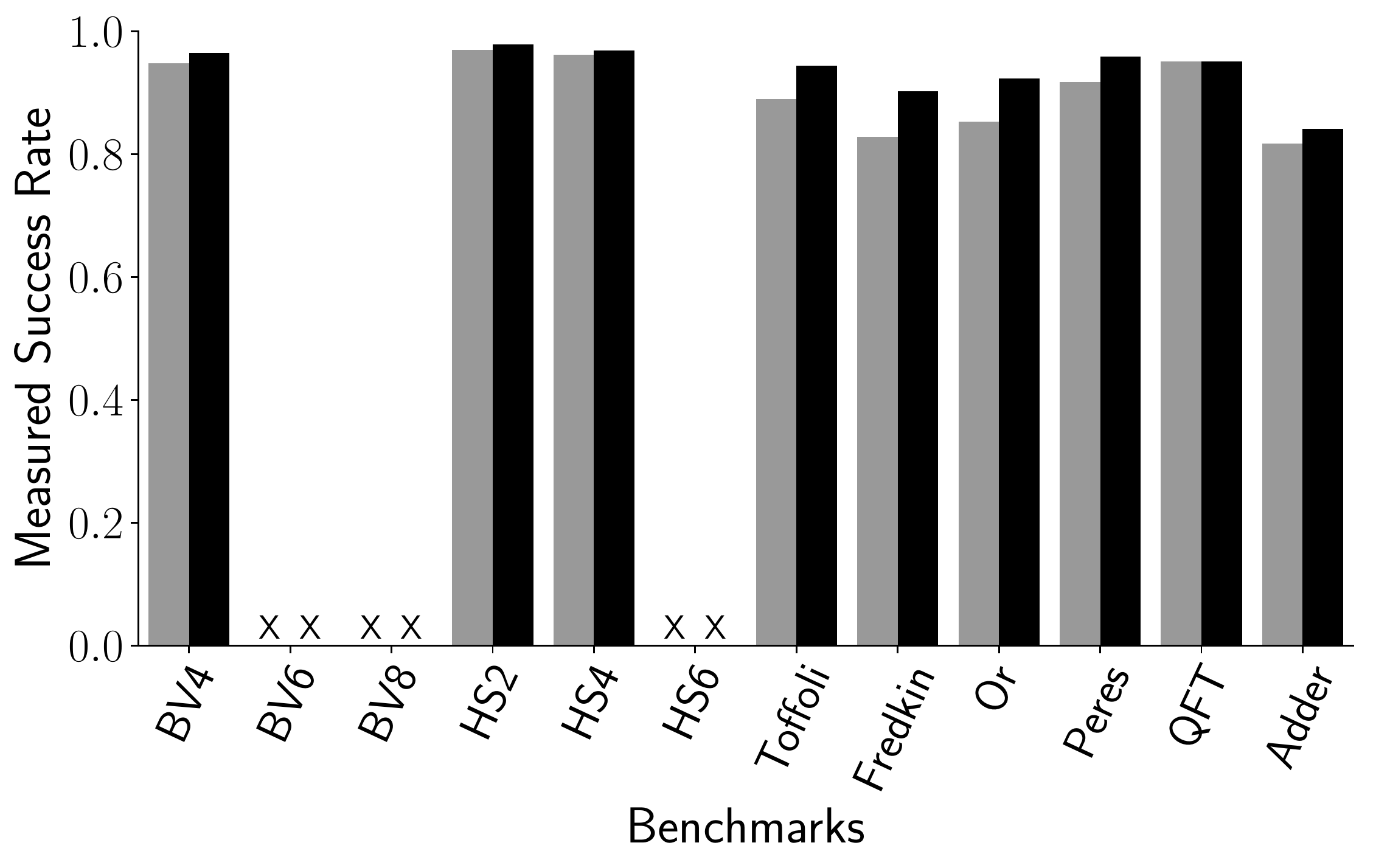}
    \label{fig:e1_gs_tion_succ}
    }
    \caption{\textbf{Importance of single qubit gate optimization.} Measured success rate for \noopt and \oneqopt. Across machines, \oneqopt coalesces operations and maximizes the use of error-free Z rotations, providing up to 1.26x improvement in success rate. In (a), the bars with zero height correspond to failed runs where the correct answer did not dominate in the output distribution. In (b), HS6, BV6 and BV8 are marked ``X'' because of size restrictions on UMDTI.}
    \label{fig:e1_gs_succ}
\end{figure}

Figure \ref{fig:e1_gs_succ} shows how 1Q optimizations result in improved success rates for IBMQ14 and UMDTI\footnote{The Rigetti machines were not available at the time these were gathered.}. Even though 1Q operations are lower-error than 2Q operations, reducing the number of 1Q operations reduces faulty operations and increases success rate substantially. 

1Q optimizations are clearly important, and the leverage gained from them depends partly on the native gate set provided by the vendor.  Across the 3 machines, \oneqopt compared to \noopt offers geomean 1.4x improvement in operation count on IBMQ14, 1.4x on Rigetti and 1.6x on UMDTI. These improvements come from mapping more effectively onto the underlying native 1Q gates, as well as by exploiting the error-free Z-axis rotations.  These 1Q optimizations pay off in success rate improvements: up to 1.26x improvement in success rate (geomean 1.09x on IBM, 1.03x on UMDTI) compared to \noopt. 

The higher gains that UMDTI sees are directly related to the gate set provided.  Namely,  the underlying hardware supports a arbitrary $R_{x,y}(\theta, \phi)$ 1Q rotation gate, which it makes software-visible. Using this operation the compiler can simplify a long sequence of 1Q gates into a single rotation operation.  This demonstrates the power of appropriate software-visible gates that can be implemented efficiently and at low error rates on the underlying hardware.

\begin{figure*}[t]
    \centering
    \subfloat[2Q Gate Counts for IBMQ14]
    {
    \includegraphics[scale=0.25]{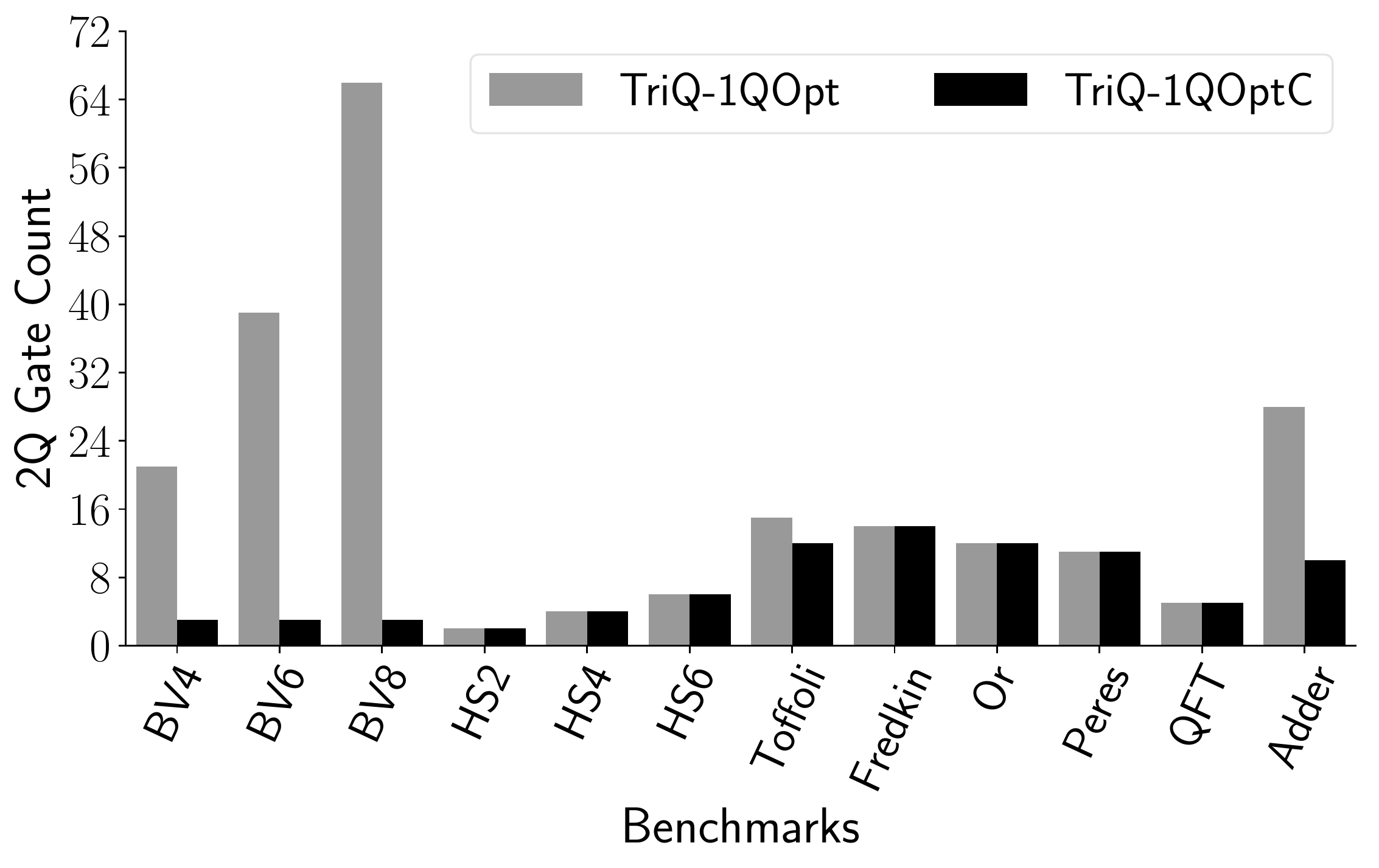}
    \label{fig:e2_comm_qmel_cnt}
    }
    \subfloat[2Q Gate Counts for Rigetti Agave]
    {
    \includegraphics[scale=0.25]{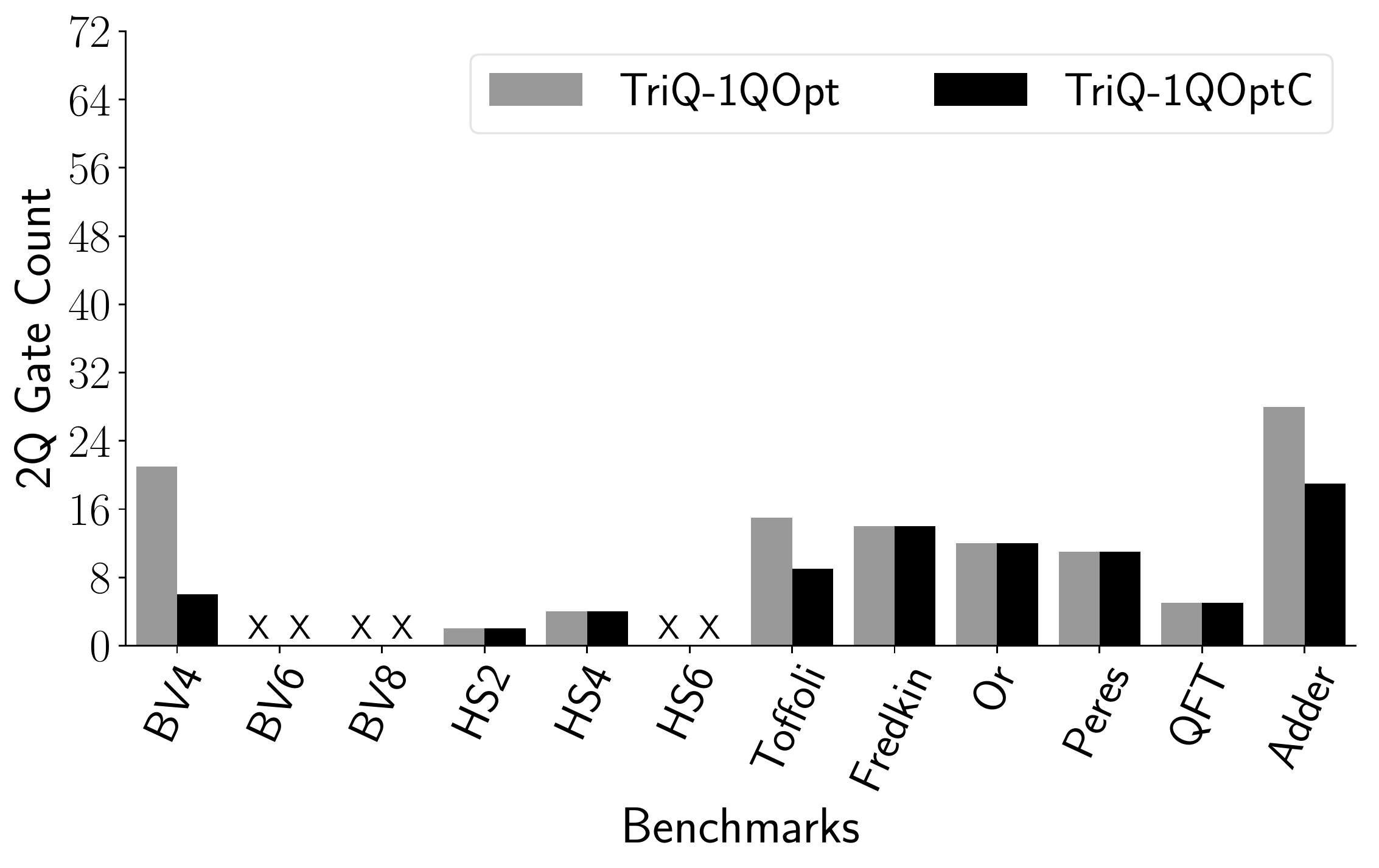}
    \label{fig:e2_comm_rigetti_cnt}
    }
    \subfloat[Success Rates for IBMQ14]
    {
    \includegraphics[scale=0.25]{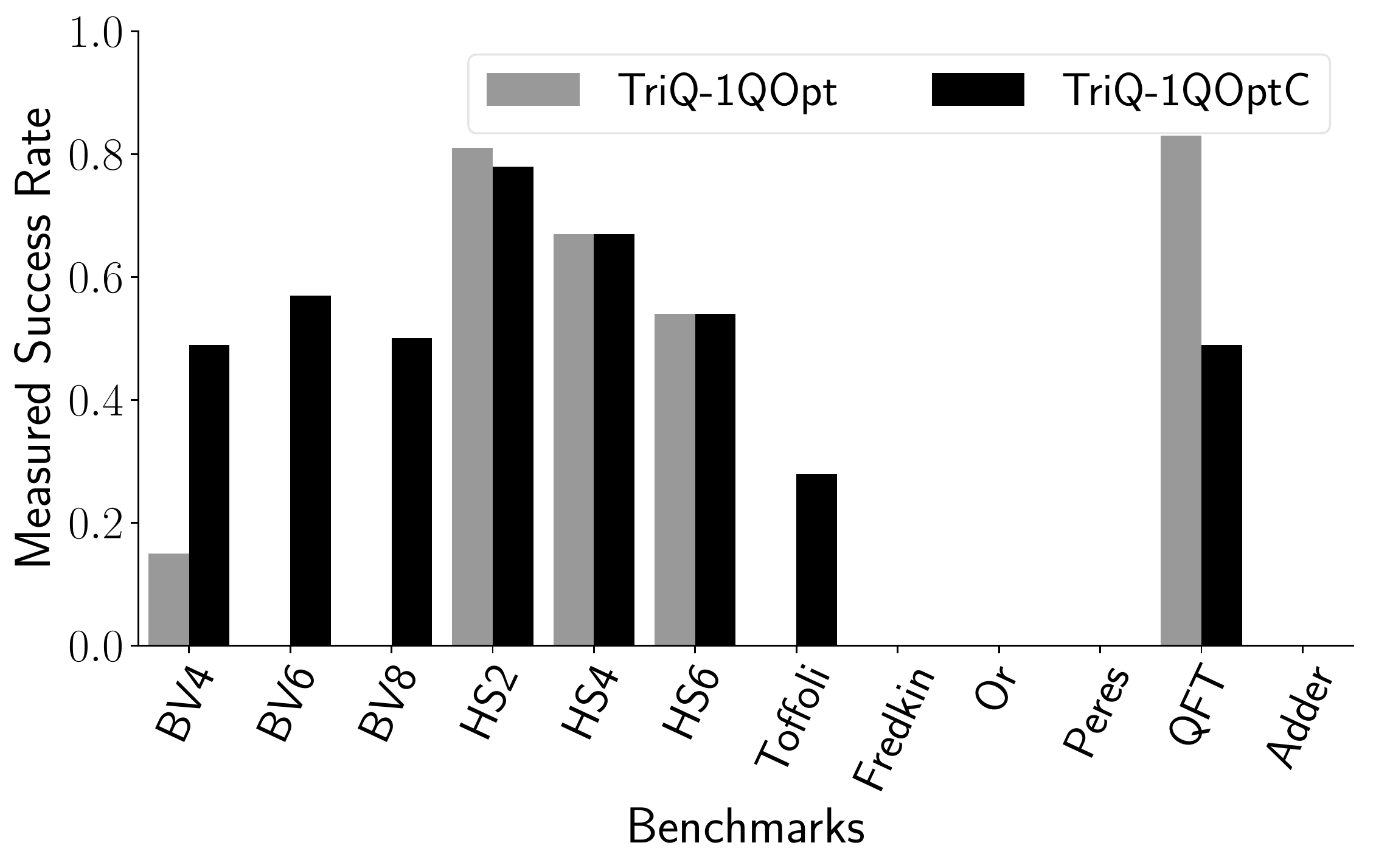}
    \label{fig:e2_comm_qmel_succ}
    }
    \caption{\textbf{Importance of communication optimization.} (a) and (b) compare the 2Q gate counts in IBMQ14 and Rigetti without and with communication optimization. (c) provides corresponding IBMQ14 success rates. On IBMQ14, \compilername provides up to 22x reduction in 2Q gate count and enables more programs to succeed compared to the default mapping. On Rigetti, \compilername obtains up to 3.5x improvement in 2Q gate count. On some benchmarks such as QFT on IBMQ14, noise-unaware communication optimization places compute on high-error resources, which leads to low success rate. In (b), BV6, BV8 and HS6 are marked ``X'' because of size restrictions on Agave. In (c) runs with zero height bars correspond to failed runs where the correct answer did not dominate in the output distribution.}
    \label{fig:e2_comm}
\end{figure*}

\subsection{Importance of Qubit Connectivity}
Figure \ref{fig:e2_comm_qmel_cnt} and \ref{fig:e2_comm_rigetti_cnt} show the 2Q gate counts for IBMQ14 and Rigetti Agave, for \oneqopt and the communication-optimized \commopt.  Where 1Q optimizations are local, 2Q optimizations are dominated by improvements in mapping the program onto communication paths that are either shorter or higher-reliability or both.   \commopt is noise-unaware, but optimizes communication by tailoring the executable to the device topology. It reduces the number of 2Q operations by up to 22x in IBMQ14 (geomean 2.1x) and up to 3.5x reduction in Rigetti Agave (geomean 1.3x). (Since UMDTI is fully-connected, these topology optimizations are not applicable here.) 

Comparing corresponding applications from Figure \ref{fig:e2_comm_qmel_cnt} and \ref{fig:e2_comm_rigetti_cnt}, Rigetti Agave (line topology) requires more 2Q operations than IBMQ14 (grid topology) because the topology is more restricted.  Toolflow functionality, i.e. \commopt, can overcome this to some degree by finding good placements, but not entirely. 

Figure \ref{fig:e2_comm_qmel_succ} shows how communication optimizations can be parlayed into higher success rate, particularly for IBMQ14. \commopt enables programs such as BV6, BV8 and Toffoli to succeed while they fail with \oneqopt. The figure  also shows the importance of program-machine topology match. Programs with  topology well-matched to the device topology have more chances of succeeding because they can be executed without swaps, reducing their 2Q gate count. For IBMQ14 (grid topology, see Figure \ref{fig:machines}), such programs include BV4 (4-qubit star) and HS2,4,6 (disjoint 2-qubit edges). On UMDTI, the fully connected topology (see Figure \ref{fig:machines}) accommodates all program 2Q patterns. Figure \ref{fig:e1_gs_tion_succ} shows that programs with varied topology have similar success rates on UMDTI.

\subsection{Importance of Noise-Adaptivity}

Noise-unaware communication optimization is useful, but not always sufficient. For example, Figure \ref{fig:e2_comm_qmel_succ} shows that for QFT, \commopt performs worse than \oneqopt. Why is that?  The machine's calibration data indicates that in doing noise-unaware mapping solely for communication distance, this compilation inadvertently resulted in qubit mappings that use less reliable hardware.  This motivates the \noiseopt approach.

\begin{figure*}[t]
    \centering
    \subfloat[2Q Gate Count on IBMQ14]{ 
        \includegraphics[scale=0.25]{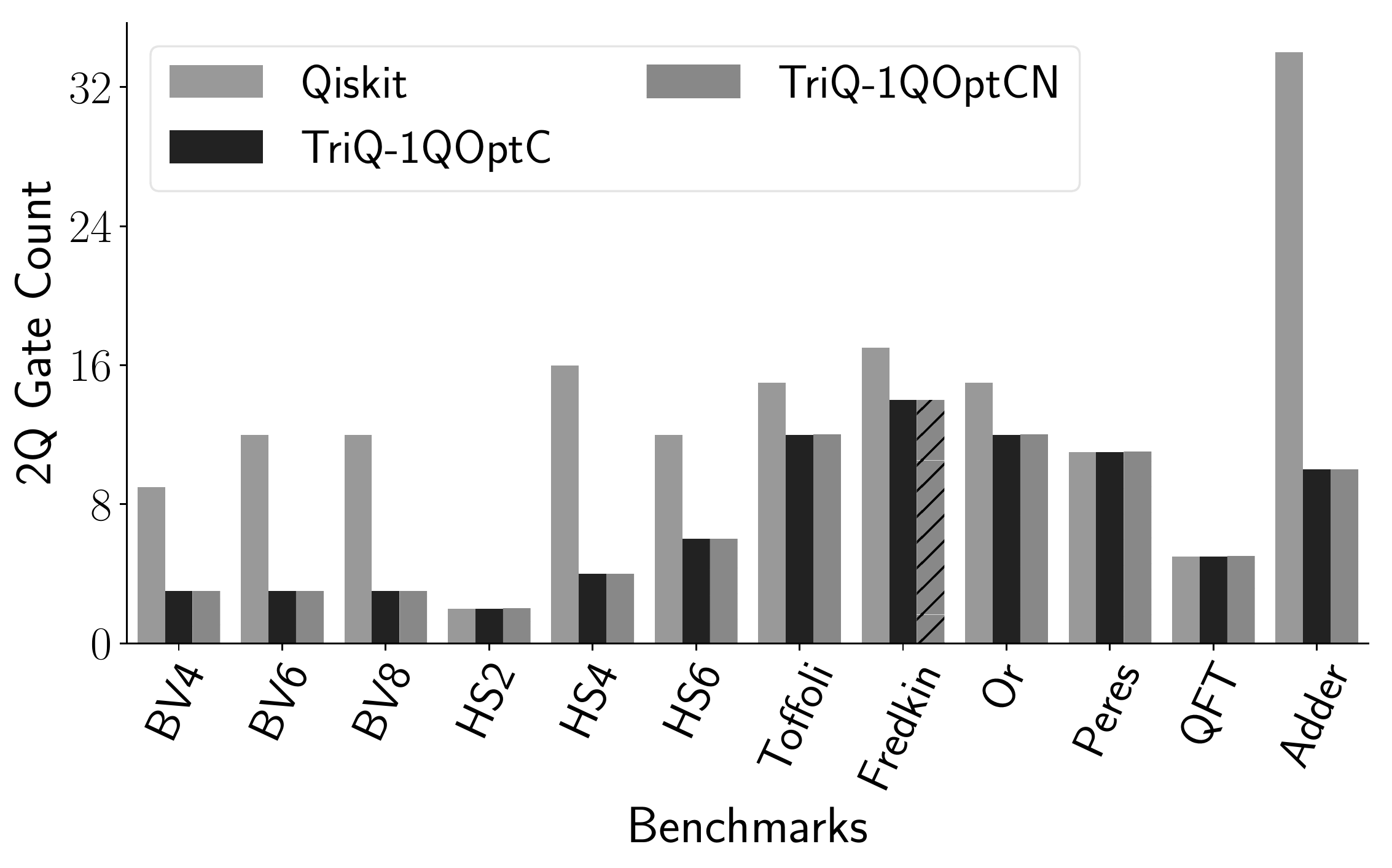}
        \label{fig:e3_noise_ibmq14_twoq}
    }
    \subfloat[Success rate on IBMQ14]{ 
    \includegraphics[scale=0.25]{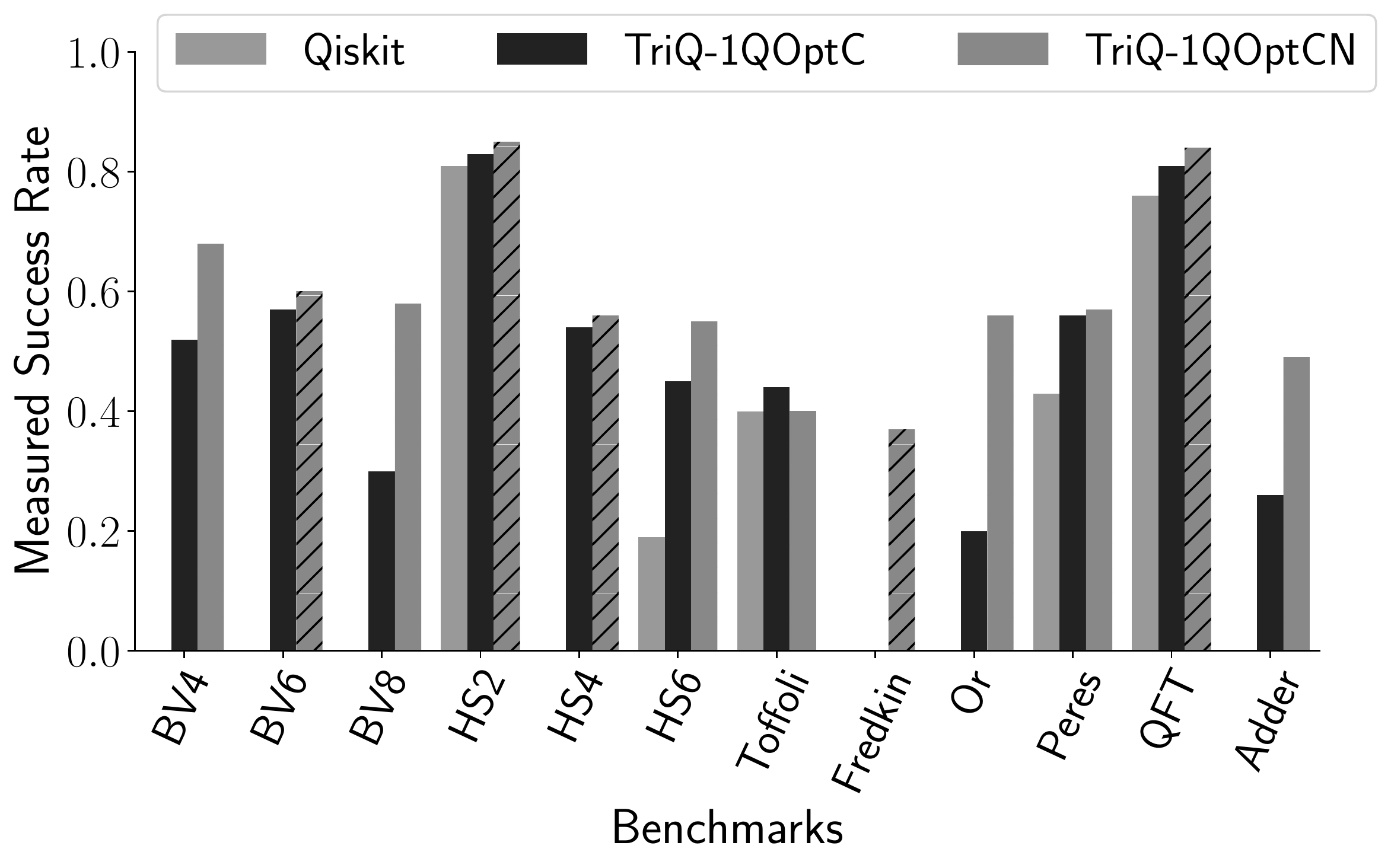}
    \label{fig:e3_noise_ibmq14_succ}
    }
    \subfloat[Success rate on Rigetti Agave]
    {
    \includegraphics[scale=0.25]{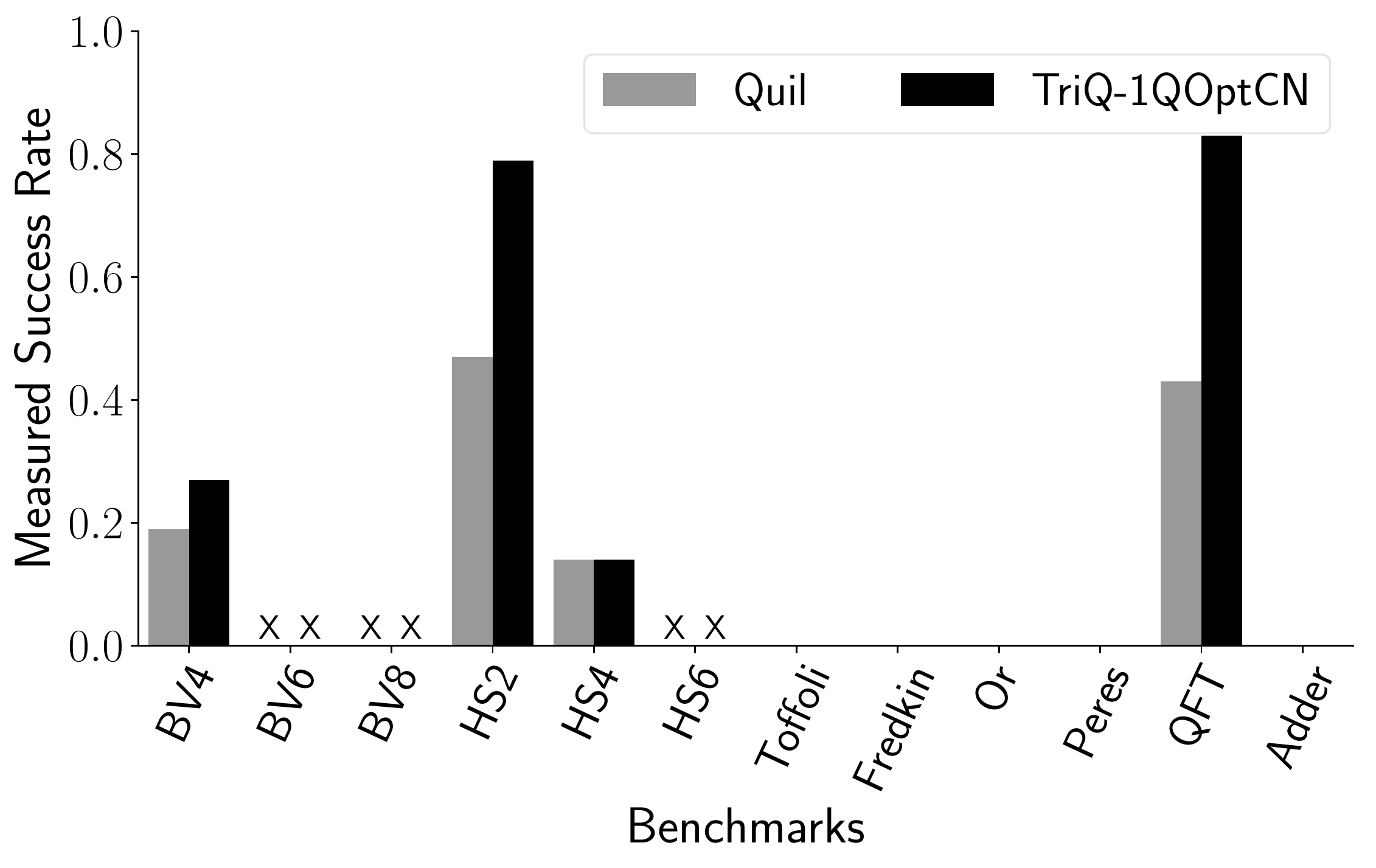}
    \label{fig:e3_noise_rigetti_succ}
    }
    
    \subfloat[Success rate on Rigetti Aspen1]
    {
    \includegraphics[scale=0.25]{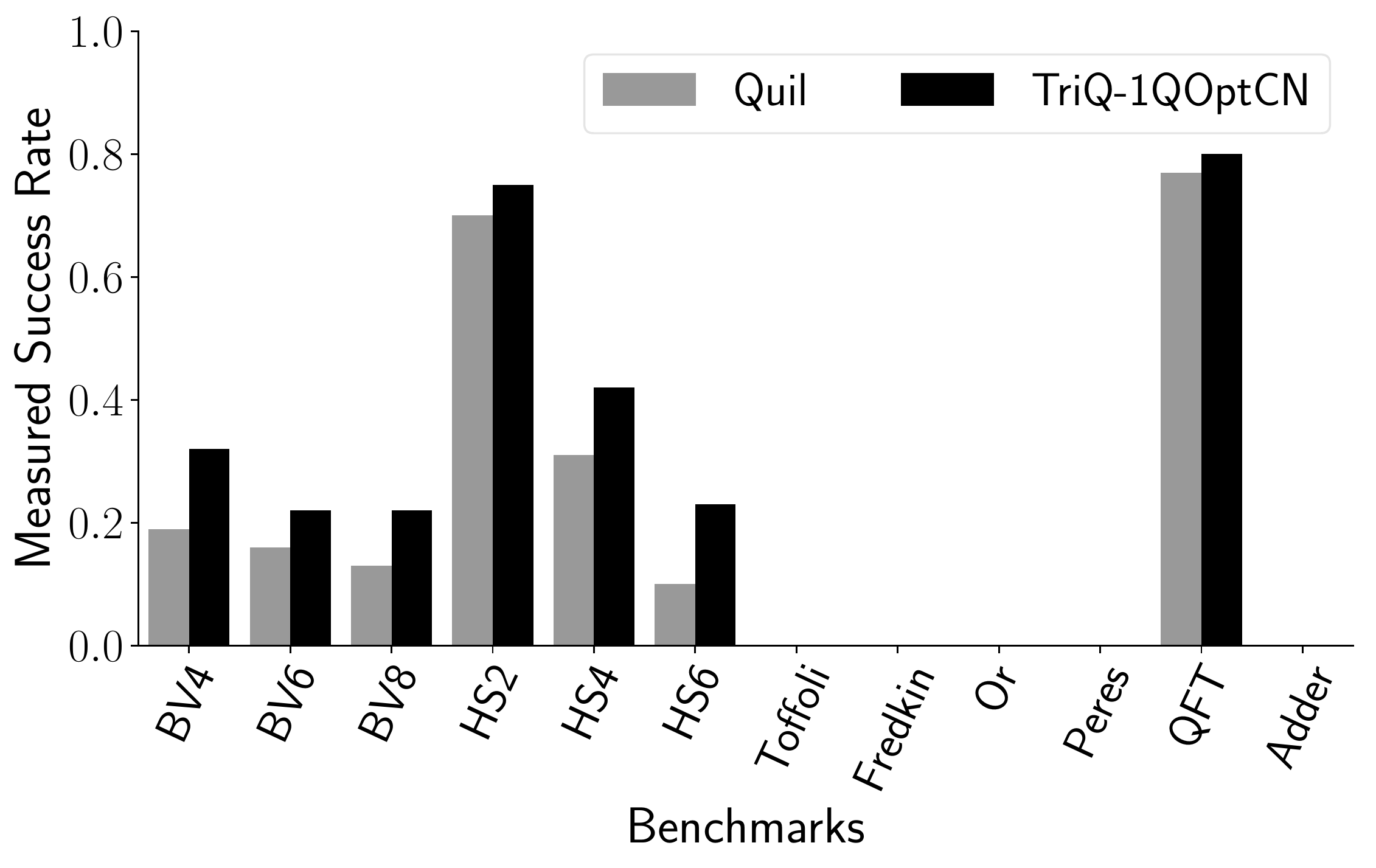}
    \label{fig:e3_noise_rigetti_aspen1_succ}
    }    
    \subfloat[Success rate for Toffoli Sequence on UMDTI]
    {
    \includegraphics[scale=0.25]{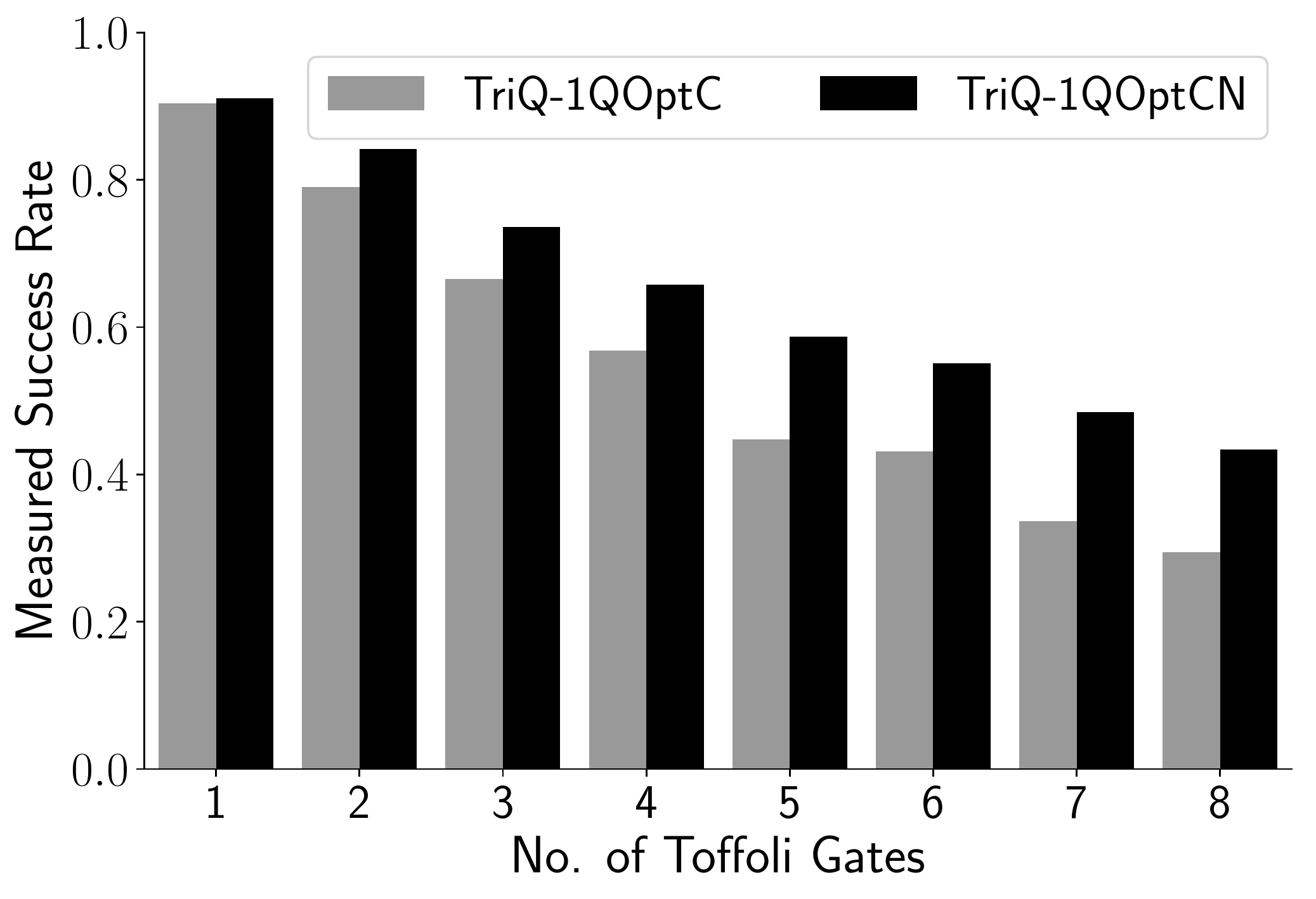}
    \label{fig:e3_noise_tion_tof}
    }
    \subfloat[Success rate for Fredkin Sequence on UMDTI]
    {
    \includegraphics[scale=0.25]{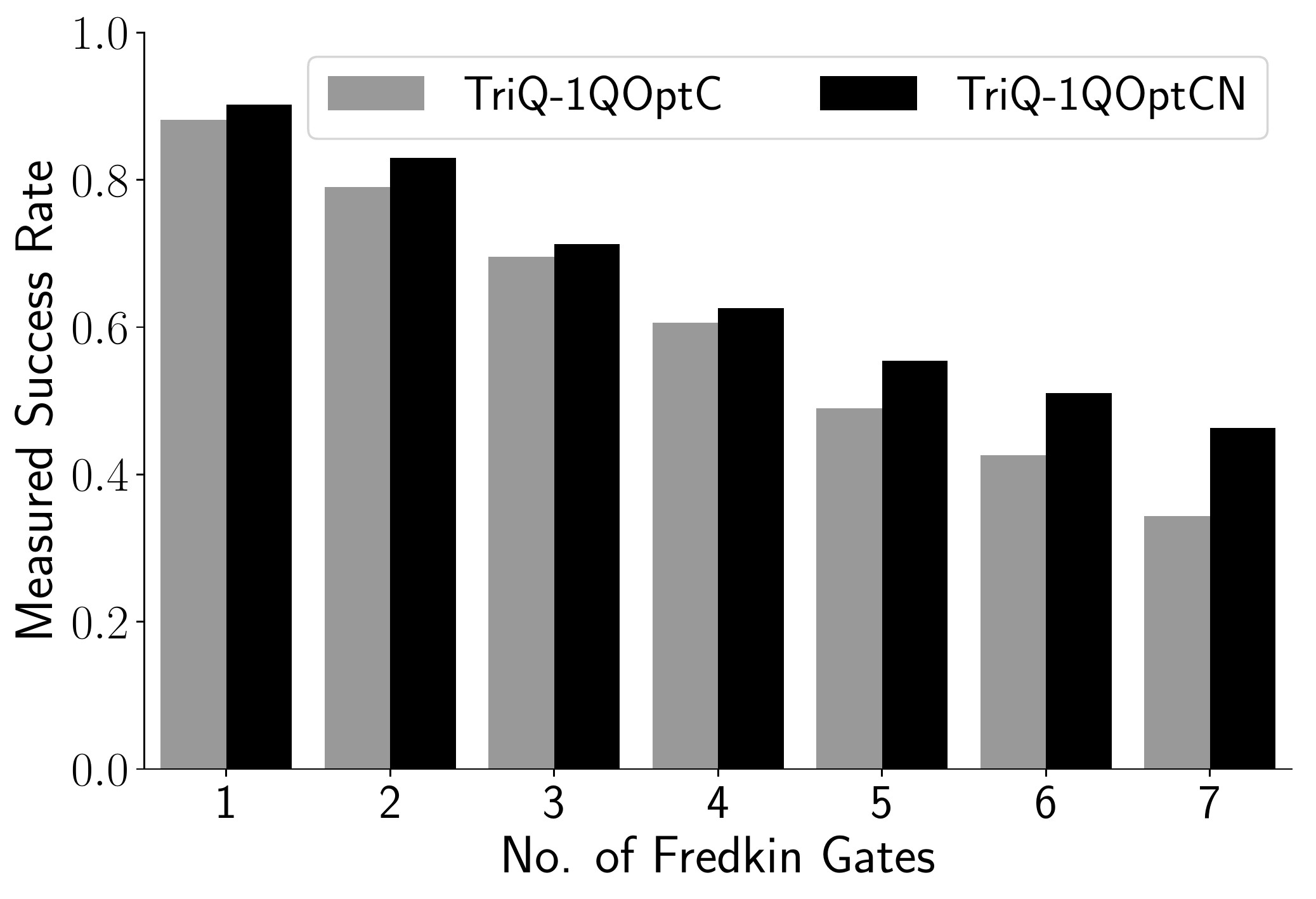}
    \label{fig:e3_noise_tion_fredkin}
    }
    \caption{\textbf{Importance of noise-adaptivity.} (a) and (b) compare \commopt, \noiseopt and IBM Qiskit compiler for IBMQ14. By optimizing gate errors, communication and single qubit gates simultaneously, \noiseopt obtains up to 28x improvement over Qiskit and 2.8x over \commopt. (c) and (d) compare Rigetti's Quil compiler and \noiseopt on Rigetti Agave and Aspen1. \noiseopt obtains up to 2.3x improvement over Quil. (e) and (f) compare \commopt and \noiseopt on UMDTI, where noise-adaptivity provides up to $1.47$x improvement. In (b), (c) and (d) runs with zero height bars correspond to failed runs where the correct answer did not dominate in the output distribution.}
    \label{fig:e3_noise_results}
\end{figure*}

Figure \ref{fig:e3_noise_results} shows the success rate and 2Q gate count for \commopt, \noiseopt, and \Qiskit. The large reductions in 2Q gate counts because of communication optimization give our methods significant benefits. 
Qiskit uses lexicographic mapping of qubits and performs swap optimization uses a greedy stochastic algorithm. It underperforms our methods because it always uses the first few qubits in the device regardless of noise and program communication requirements. 2Q optimizations again parlay into success rate benefits.
\noiseopt succeeds on all 12 benchmarks, and outperforms \commopt by up to 2.8x (geomean 1.4x). In Toffoli, it under performs slightly, but the loss is comparable to the noise-margin in these runs and not significant. \commopt fails to produce the correct answer in the Fredkin benchmark, while \noiseopt succeeds. In contrast, the IBM Qiskit compiler (noise-unaware) fails to produce the correct answer on 7/12 benchmarks. To compute improvement factors, we used the measured probability of the correct answer produced by Qiskit (other incorrect answers had higher probability). \noiseopt obtains up to 28x improvement (geomean 3.0X) over Qiskit.

Figure \ref{fig:e3_noise_rigetti_succ} and \ref{fig:e3_noise_rigetti_aspen1_succ} show how \noiseopt can improve in success rate on Rigetti. For most benchmarks we see significant improvements. We obtain up to 2.3x improvement over the Quil compiler (geomean 1.45x). Quil uses a simple initial qubit mapping, with insufficient communication optimization and no noise-awareness.  

Finally, Figure \ref{fig:e3_noise_tion_tof} and \ref{fig:e3_noise_tion_fredkin} show the success rate improvements of \noiseopt for UMDTI. This machine has a fully-connected topology and low error rates. For the applications that fit into its 5 qubits, success rates are high across the board; therefore, we created more challenging 3-qubit benchmarks with longer gate sequences.  Namely, we used iterations of Toffoli or Fredkin gate sequences to test the effects of noise on programs with varying lengths.
\noiseopt offers 1.47x improvement for Toffoli and 1.35x improvement for Fredkin sequences, compared to \commopt. The gains increase with increasing program length. \noiseopt obtains these improvements because it places frequently interacting pairs of program qubits on the best hardware. 

Whereas superconducting qubits can have fundamental physical defects \cite{superconducting_stability}, each qubit in an ion trap identical and defect-free. However, noise in ion trap machines stems from difficulties in qubit control and sensitivity to motional mode drifts; these cause 1-3\% fluctuations in error rates.
Our UMDTI experiment is the first to demonstrate that longer programs can obtain performance improvements by adapting to these small noise variations. For larger ion traps, reduced interaction strengths and therefore higher error rates are expected between ions which are farther apart \cite{iontrap14q, iontrap_blueprint}. This suggests that our noise-adaptive methods will be even more important then.

\subsection{Putting it all together}
We used \compilername to compile the 12 benchmarks for all the \syscnt systems using all the optimizations i.e., \noiseopt. Since \noiseopt outperforms vendor compilers, it allows us to understand benchmark performance without introducing compiler inefficiencies.

\begin{figure*}[t]
    \centering
    \includegraphics[scale=0.26]{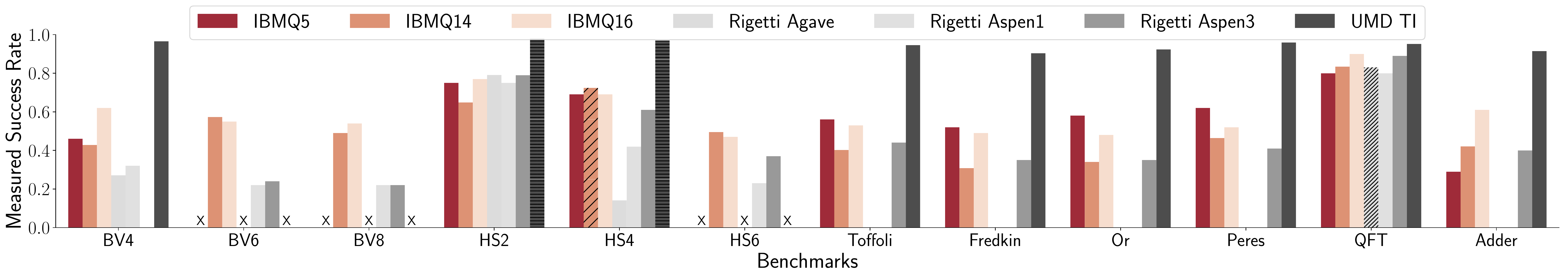}
    \caption{Success rate for 12 benchmarks on \syscnt systems. Success rates varies drastically across systems and is influenced by error rates, qubit connectivity, and application-machine topology match.  Benchmarks that are too large to be mapped onto a machine are marked ``X''. Runs with zero height bars correspond to failed runs where the
correct answer did not dominate in the output distribution. \textit{This comparison is intended to understand the impact of architectural design choices such as gate set and connectivity on benchmark performance and is not intended to pick a winning technology, vendor or implementation. Individual benchmark performance numbers may change over time --- these measurements represent a snapshot of the performance of these systems when we performed the experiments. }}
    \label{fig:bench_perf}
\end{figure*}

Figure \ref{fig:bench_perf} shows the success rate measured for each of the compiled executables. On benchmarks that fit on the current UMDTI machine, its low gate errors and good topology give it an advantage over other machines. For the superconducting machines, application-device topology match matters. For example, the Toffoli, Fredkin and Or benchmarks (3-qubit triangle) fit the triangular topology of IBMQ5 (see Figure \ref{fig:machines}). This offers higher performance than the grid topology of IBMQ16. Assuming reasonable application-topology match, having more qubits is better because it allows more flexibility in choosing a mapping which avoids the noisy regions of the machine. 

Rigetti's best gates are comparable to IBM (IBMQ5 and Rigetti on HS2 and QFT).  Application-topology mismatch and higher noise variation affect the success rate of larger benchmarks. Comparing the results on the newer Aspen1 and Aspen3 machines and the older Agave machine indicates significant improvements in qubit and gate reliability. On Aspen1 and Aspen3, more powerful native operations can be exploited to reduce the number of 2Q operations for some of our benchmarks. These operations were not software-visible on Aspen1 and Aspen3 in our experiments; exposing them to the compiler would enable higher success rates. 

\subsection{Scalability of our Toolflow}
Finally, although compile time has not been a prominent goal for this work, we wanted to test the degree to which our optimization approaches would scale up to larger NISQ machines that do not yet exist.  Toward this goal, we measured the toolflow runtime to perform full optimizations (including noise-awareness) for proposed Quantum Supremacy circuits, mapping onto the  72-qubit system announced by Google \cite{googlebristlecone} (but not yet operational). We assigned error rates for each gate by sampling from 100 days of error data from IBM devices.  \noiseopt scales well up to 72 qubits and is three orders of magnitude faster than \cite{asplos}. The scaling is independent of gate count because the solver only creates variables for distinct two-qubit gates between a set of $n$ program-qubits, bounding the number of variables by $O$($n^2$).  This evaluation gives us confidence that our approaches will scale through NISQ machines of considerable size and capability.
\section{Architecture Implications}
\label{sec:archimplics}

Section \ref{sec:results}'s results have already offered important insights regarding the interplay of hardware design choices (e.g. qubit technologies, communication topologies etc.) and the software toolflows and execution stack.  In particular, QC sits at an exciting inflection point where decisions about the hardware-software interface are timely and important.  

{\noindent \textbf{Native gates and software-visible operations}:} Our experiments show that when native gates are software-visible, the compiler can optimize programs heavily, in ways that are well-suited to the underlying device often resulting in lower error rates.  This provides both execution time and success rate benefits. Ideally, vendors should expose the most fundamental 1Q and 2Q operations, allowing the compiler to perform fine-grained optimization across many gates. Such optimizations are especially relevant in systems such as UMDTI and Rigetti, where several native gates are required to construct a single CNOT gate. Providing powerful native operations such as UMDTI's flexible 1Q rotations provides further benefits, enabling a large number of operations to be simplified into fewer operations on the qubits. From a technology perspective, it may be difficult for some qubit technologies to support very flexible native operations.  When such limits are reached, our results show the value of exposing to software what native operations {\em are} available, to allow ample compiler optimization.  These observations are also corroborated by IBM's recent announcement that they will expose pulse-level qubit control via a programmable interface \cite{openpulse_arxiv}. As an analogy to classical microprocessors, this is akin to making micro-operations software-visible \cite{Wilkes:1989:BWD:94938.94976}.
 
 {\noindent \textbf{Communication topology}:} 
 Our experiments show that communication topology has a significant impact on performance and success rate. Comparing near-neighbor vs. fully-connected implementations like IBMQ14 vs. UMDTI, our results show that machines with richer qubit connectivity allow a wider variety of programs to execute successfully.  Our results also demonstrate that when full connectivity cannot be reached (e.g. offering it is physically easier in ion traps, compared to superconducting systems) compilers like ours can optimize communication for the achievable topology.
 
 {\noindent \textbf{Noise rates and variability}:} 
 Clearly, lower error rates are preferable---crucial for benchmark success rate.  Nonetheless, our studies show the degree to which noise-aware toolflows can be beneficial even on low-error platforms. Even small variations in low error rates like UMDTI's have significant impact on application success. Comparing devices such as Rigetti Aspen3, IBMQ16 and IBMQ14, it is also important that large connected segments of the system have good error rates; contiguous regions of low-error gates are the most useful. 
 Finally, it is already the norm in QC to compile programs for a particular input size, and our work further demonstrates the value of also  recompiling applications to account for up-to-date noise data as well.
 
{\noindent \textbf{Execution stack for QC}:} 
The \compilername toolflow has core functionality which is portable across diverse platforms, but also allows full top-to-bottom optimizations for device and application characteristics provided as compile-time inputs. Our results show the value of these device-aware and application-aware characteristics in exploiting successfully the limited resources of NISQ machines.  Exploiting low-error hardware and native gates was crucial to our success rates. This points to the conclusion that QC machines may not yet be ready for abstractions or virtualizations that abstract too much of the information flow between software and hardware. 
\nocite{koen_bertels1, koen_bertels2, koen_bertels3} 

\nocite{qiskit, openqasm1} 
\nocite{quil, pyquil} 



\section{Related Work}

With prototype QCs quite recent, \cite{bench2} provided the first experimental comparison on IBMQ5 and UMDTI using 4 hand-optimized benchmarks on 2 5-qubit machines, and observed the importance of program-device topology match. Our work extends to more and considerably larger prototypes.  We are also the first to do multi-platform characterizations via a a top-to-bottom compiler toolflow for real-system executions across platforms. Where \cite{bench2} hand-placed each qubit, our work is the first to perform cross-platform comparisons of optimizations leveraging qubit, topology and noise properties within automated multi-platform compilers. These cross-platform optimizations and observations inform Section \ref{sec:archimplics}'s summary of the work's hardware and architectural implications.

At the other end of the stack, QC programming languages and compilers have seen ongoing attention.  In addition to Scaffold, other examples are Quipper \cite{quipper1, quipper2}, and LIQUi$\ket{}$ \cite{liquid1}. IBM Qiskit is a Python-based framework to program and compile code for the IBM systems, generating OpenQASM \cite{openqasm1}. 
Likewise PyQuil \cite{pyquil, forest} is a Python-based framework for Rigetti systems, generating Quil \cite{quil}. ProjectQ \cite{projectq1, projectq2} is another Python-based framework to describe quantum circuits and compile them for different machines. 
Qiskit and Quil consider communication optimization while ProjectQ does not currently support non-grid topology \cite{projectq_bug}.  Through its top-to-bottom empirical results, our work shows the importance of both device-specific and application-specific optimizations through the toolflow. We demonstrate how to achieve this with a core set of passes that apply across diverse platforms, by taking device characteristics as input.

\cite{intel1, ai1, cgo18, zulehner1, zhang18} develop methods for optimizing communication on current or small systems, but do not consider noise data. Compared to the open source implementation of \cite{zulehner1}, TriQ reduces 2Q gate count by 1.2x (geomean), up to 2X. In \cite{tannu_qureshi}, they  propose the use of noise-aware qubit mapping and movement policies on the 20-qubit IBM system and report real executions on IBMQ5. For BV4, \cite{tannu_qureshi} reports a success rate of 0.23 on the 5-qubit IBM system. Since the machine state influences success rate, in order to make a fair comparison, we evaluated TriQ on 6 days having different error conditions. We obtained 2X better success rate ranging from 0.43 to 0.51 (average 0.47) indicating that our optimizations are effective. \cite{asplos} developed a noise-aware compiler for the Scaffold language, targeted for systems with grid topologies and demonstrated the benefits of noise-adaptive compilation on IBMQ16. However, none of these provided multi-platform optimizations. Ours is the first work to build a full-stack toolflow in support of cross-platform empirical experiments comparing multiple QC prototypes.

\section{Conclusions}
After decades of gradual progress, NISQ QC prototypes are now available for experiments. Several machines exist in the 5-50 qubit range, representing  widely-divergent design points regarding qubit technologies, topologies, and error rates.  This diversity offers opportunities for cross-platform design studies that elucidate how device technologies influence other hardware design choices, and how compiler and software choices can offer optimizations that mitigate challenging hardware characteristics.   To study key system design questions, 
our work built \compilername, a top-to-bottom toolflow which compiles high-level language programs for multiple target systems.  Using real-systems measurements on \syscnt devices, our experiments with \compilername show several examples of how leveraging hardware details in the compiler can provide a significant boost in program success rates. Our empirical cross-platform and cross-technology study offers forward-looking insights for compiler and architecture design for NISQ systems.

\begin{acks}
This work is funded in part by EPiQC, an NSF Expedition in
Computing, under grants CCF-1730082. We thank Christopher Monroe, Kevin Akiva Landsman and Daiwei Zhu from University of Maryland for access to the ion trap system. We thank Ryan Karle, Marcus da Silva, Amy Brown, Tushar Mittal and Nima Alidoust from Rigetti. We thank Ken Brown and Fred Chong for their insightful comments and suggestions. C.H.A. acknowledges financial support from CONACYT doctoral grant number 455378.
\end{acks}
\bibliographystyle{ACM-Reference-Format}


\end{document}